\documentclass[Onecolumn,12pt %
 reprint,
superscriptaddress,
 amsmath,amssymb,
]{revtex4-2}
\usepackage{float}
\usepackage{graphicx}
\usepackage{dcolumn}
\usepackage{bm}

\usepackage{times} 

\newcommand{\bn}{\begin{align}}
\newcommand{\enl}[1]{\label{#1}\end{align}}

\newcommand{\ba}{\begin{equation}}
\newcommand{\bs}{\begin{split}}
\newcommand{\ea}{\end{equation}}
\newcommand{\el}[1]{\label{#1}\end{equation}}
\newcommand{\bsa}{\begin{subequations}}
\newcommand{\esa}{\end{subequations}}
\newcommand{\esl}[1]{\label{#1}\end{subequations}}
\newcommand{\baq}{\begin{eqnarray}}
\newcommand{\eaq}{\end{eqnarray}}

\newcommand{\ket}{\rangle}

\usepackage{mathptmx}
\usepackage{xcolor}
\usepackage{setspace}
\usepackage{mhchem}
\usepackage{subcaption}
\begin{document}

\preprint{APS/123-QED}

\title{Dual frequency calibration to build a portable vapor cell optical clock with improved stability  and without a frequency comb}

\author{Sreeshna Subhash}\author{Greeshma Gopinath} \author{Sankar Davuluri}\email{sankar@hyderabad.bits-pilani.ac.in}
\affiliation{%
 Department of Physics, Birla Institute of Technology and Science, Pilani, Hyderabad Campus, Jawahar Nagar, Kapra Mandal, Medchal District, Telangana 500078, India
}%


\date{\today}
\setstretch{1.2}
\begin{abstract}
This article theoretically proposes a new dual interferometer technique to accurately calibrate two laser frequencies simultaneously using four-wave mixing in an alkali metal vapor cell. The two frequency-calibrated lasers are mixed to create a beat signal at radio frequency to build a portable optical atomic clock (OAC) without an optical frequency comb (OFC). Removal of the OFC improves the portability of OAC, while the dual interferometer setup enhances the one second stability to $1.3\times 10^{-15}$, which is better than the current portable OAC. Thermal noise in the OAC is minimized by choosing the double-lambda atomic scheme with co-propagating laser fields. Using D2 transition of $\ce{^{87}_{}Rb}$, the standard quantum limited frequency sensitivity and stability of the OAC are estimated as $3.2\;\sqrt{\mbox{Hz}}$, and $1.3\times10^{-15}\sqrt{\mbox{Hz}^{-1}}$, respectively. After considering broadening effects due to $357\,$K temperature and collisions, the optimum stability of the OAC is reduced to $3.3\times10^{-15}\sqrt{\mbox{Hz}^{-1}}$ for a laser with $1\,$KHz linewidth and $0.54$ mW power at the input of the vapor cell.
\end{abstract}

\maketitle
\section{Introduction}
An optical atomic clock (OAC) \cite{RevModPhys.87.637,Poli2013,Roslund2024,bandi2024comprehensive} is the most advanced timekeeping device with stability up to $10^{-17}$ at one second averaging time \cite{Schioppo2017,Bothwell_2019,McGrew2018,PhysRevLett.123.033201}. Achieving such high stability requires relatively complex setups \cite{PhysRevLett.109.180801,PhysRevLett.126.011102,PhysRevLett.128.033202,Zheng2022,lasing,Ushijima2015} with optical trapping and cryogenic cooling. A second category of OAC, which is of immense interest in space applications \cite{doi:10.1126/science.1192720}, is portable OAC \cite{Newman:19}. Portable OAC offers a less complex design at the cost of reduced stability compared to complex ion clocks \cite{PhysRevLett.134.023201,PhysRevLett.104.070802,PhysRevApplied.17.034041,PhysRevApplied.9.014019}. The best example of portable OAC is vapor cell clocks \cite{Shi:24,PhysRevApplied.21.024003,https://doi.org/10.1002/navi.215,Newman:21}, most of them currently exhibit a stability of $10^{-12}$ to $10^{-13}$ at the $1\,$ second averaging time.

The advanced portable OAC is based on two-photon absorption in three-level atoms \cite{PhysRevApplied.9.014019}. An intense laser field excites an atom from its ground state to its highest level through two-photon absorption in a three-level cascade scheme \cite{Newman:21}. The excited atom de-excites to the ground state through fluorescence from an auxiliary level \cite{Callejo:25}. A maximum amount of fluorescence implies that the frequency of the incident laser is on two-photon resonance with the
atoms. Two-photon absorption requires a strong laser field intensity, which shifts the atom's energy levels through the AC Stark effect \cite{PhysRevA.100.023417,PhysRevApplied.14.024001}. The AC Stark shifts can be minimized by using non-degenerate lasers for two-photon absorption, but this trick increases the thermal broadening \cite{PhysRevApplied.10.014031}. Minimizing thermal broadening in the cascade scheme requires the use of counterpropagating lasers with a degenerate frequency \cite{Hafiz:16}. Hence, a trade-off between the Stark effect and thermal broadening must be achieved to optimize the clock's sensitivity. The stability of these clocks has reached \cite{s22051982} $10^{-13}$ at 1 s averaging time. 

Coherent population trapping (CPT) clocks are another prominent design using vapor cells \cite{PhysRevA.108.013103,10.1063/1.4931768}. The CPT scheme uses three-level atoms in lambda geometry \cite{Yudin:17,Fang2021}. Several detection schemes \cite{PhysRevLett.79.3865,PhysRevApplied.12.054063} have been developed using the fact that the fields have maximum transparency when the two-photon resonance condition is satisfied. A prominent method \cite{PhysRevA.108.013103} is to modulate a laser field frequency using a microwave field. When the microwave field frequency equals half the frequency difference of the lower two hyperfine levels in the lambda scheme, the applied optical field frequency is modulated into two frequency components that satisfy the two-photon resonance condition. 
 
Once the laser frequency is calibrated to an atomic resonance, for timekeeping, a beat signal at radio frequency is created by mixing the laser with a known tooth of optical frequency comb (OFC) \cite{970894,Fortier2019}. This article focuses on improving the stability and portability of vapor cell OAC by eliminating the requirement of a OFC.
 The OFC is eliminated by calibrating two laser frequencies simultaneously, and then using them to create a beat signal at radio frequency for timekeeping. A novel dual interferometer setup with four-wave mixing \cite{PhysRevA.83.063823} is proposed to calibrate two laser frequencies.
The paper is organized as follows. Section~\ref{model} describes the atom-laser interaction required to build the OAC. Section~\ref{clock-design} describes the dual interferometer design for the simultaneous calibration of two laser frequencies. Section~\ref{broadening} describes the stability of the OAC in the presence of several broadening mechanisms. Section \ref{Discussion} is the discussion. 
\section{Model}\label{model}
\begin{figure}[hbt!]
\centering
\includegraphics[scale=1.3]{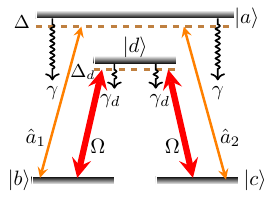}
\caption{The atomic states are $|b\rangle=|[5S_{1/2}, F=1, m_F=-1]\rangle$ and $|c\rangle=|[5S_{1/2}, F=1, m_F=1]\rangle$, $|d\rangle=|[5P_{3/2}, F=0, m_F=0]\rangle$ and $|a\rangle=|[5P_{3/2}, F=1, m_F=0]\rangle$ of rubidium ($\ce{^{87}_{}Rb}$) atom. The population decay rate from $|a\rangle\to |b\rangle$ ($|d\rangle\to |b\rangle$) and $|a\rangle\to |c\rangle$ ($|d\rangle\to |c\rangle$) is $\gamma$ ($\gamma_d$). The transitions on $|d\rangle-|b\rangle$ and $|d\rangle-|c\rangle$ are driven at Rabi frequency $2\Omega$ by classical driving lasers, while $|a\rangle-|b\rangle$ and $|a\rangle-|c\rangle$ are coupled to weak quantum fields described by annihilation operators $\hat{a}_1$ and $\hat{a}_2$, respectively.}
\label{fig1}
\end{figure}
Figure. \ref{fig1} shows an atomic scheme with energy level $|x\rangle$ $(x = a,b, c,d)$ and energy $\hbar\omega_x$, where $\hbar$ is the reduced Planck constant and $\omega_x/2\pi$ is the frequency. The transitions $|b\rangle-|a\rangle$ and $|c\rangle-|a\rangle$ are coupled to two weak quantum fields
that have mutually orthogonal circular polarizations. The propagation of each quantum field is described by using quasi
mono-chromatic annihilation operators \cite{PhysRevA.35.4661,PhysRevA.42.4102} $\Sigma_r \hat{c}_{rj}$, $j = 1,2$, with wave vector $k_{rj}$ and the frequency $\nu_{rj}/2\pi$ for the $r-$th mode of the $j$-th quantum field. The relation between $k_{rj}$ and mean wave-vector $k$ are given as
\begin{equation}\label{eq.1}
k_{rj}=k+\frac{2\pi r}{l},\quad \nu_{rj}=\nu+\frac{2\pi r c}{l},\quad r=-R,...,R,
\end{equation}
where $c$ is the velocity of light in vacuum, $\nu_{rj}=ck_{rj}$, $\nu=ck$, and $l$ is the quantization length. The quantum field interaction with $N$ number of atoms in a length $l$ can be described by using the technique described in Refs. \cite{Drummond:87,PhysRevA.49.1973}. We divide $l$ into $(2R+1)$ subcells with each subcell centered at $z_p=pl/(2R+1)$, $p=-R,...,R$, such that $\Delta z:=z_{p+1}-z_p=l/(2R+1)$. By using the notation $\hat{\sigma}_{xy}^{ps}=|x\rangle^{ps}\langle y|^{ps}$ (where $y=a,b,c,d$) with the superscripts $s$ and $p$ denoting the $s-$th atom in the sub-cell centered at $z_p$, the Hamiltonian $\hat{H}$ of the system is described as
\begin{equation}\label{h1}
\begin{split}
\frac{\hat{H}}{\hbar} &=\sum_{x=a}^{d}\omega_x\hat{\sigma}_{xx}^{ps}+\sum_r\sum_{j=1}^2\nu_{rj}\left(\hat{c}^\dagger_{rj}\hat{c}_{rj}+\frac{1}{2}\right)+\frac{\hat{H}_r}{\hbar}+\Big(\sum_{p,s,r}\tilde\Omega\hat{\sigma}_{db}^{ps}+\sum_{p,s,r}\tilde\Omega\hat{\sigma}_{dc}^{ps}
\\&+\sum_{p,s,r}g_r\hat{\sigma}_{ab}^{ps}\hat{c}_{r1}e^{ik_{r1}z_p}
+\sum_{p,s,r}g_r\hat{\sigma}_{ac}^{ps}\hat{c}_{r2}e^{ik_{r2}z_p}+H.c\Big),
\end{split}
\end{equation}
where $\hat{H}_r$ is the reservoir Hamiltonian, $H.c$ is the Hermitian conjugate. A left circular classical driving laser, with frequency $\nu_d/2\pi$ and amplitude $E_0$, drives the transition $|d\rangle-|b\rangle$ with $\tilde\Omega=-\wp_dE_0\cos(-\nu_d t)/\hbar$. A right circular driving laser, with frequency $\nu_d/2\pi$ and amplitude $E_0$, drives the transition $|d\rangle-|c\rangle$ with $\tilde\Omega$. We assume that the $|d\rangle-|b\rangle$ and $|d\rangle-|c\rangle$ transitions have the same dipole moment $\wp_d$, while the $|a\rangle-|b\rangle$ and $|a\rangle-|c\rangle$ transitions have the same dipole moment $\wp$. $g_r=-(\wp/\hbar)\sqrt{\hbar\nu_r/2\epsilon_o V_q}$ is the coupling constant between weak quantum fields and $|a\rangle-|b\rangle$, $|a\rangle-|c\rangle$ transitions with electric permittivity $\epsilon_0$ and quantization volume $V_q$. In the limit $l\rightarrow\infty$ and $\Delta z\rightarrow 0$, by using the definitions $z_p=pl/(2R+1)\rightarrow z, (2R+1)\sum_{s}\hat{\sigma}_{xy}^{ps}|_{z_p\rightarrow z}:=N\hat{\sigma}_{xy}^0$, $(2R+1)\sum_s\hat{\sigma}_{xy}^{ps}(z_p)|_{z_p\rightarrow z}:=N\hat{\sigma}_{xy}^0(z)$, $\sum_r\hat{c}_{rj}=(\sqrt{l/c})\hat{a}_j(z,t)$, and after making the rotating wave approximation by taking $\hat{\sigma}_{aa}^0=\hat{\sigma}_{aa}$, $\hat{\sigma}_{bb}^0=\hat{\sigma}_{bb}$, $\hat{\sigma}_{cc}^0=\hat{\sigma}_{cc}$, $\hat{\sigma}_{ab}^0=\hat{\sigma}_{ab}e^{i\nu t}$, $\hat{\sigma}_{ac}^0=\hat{\sigma}_{ac}e^{i\nu t}$, $\hat{\sigma}_{bc}^0=\hat{\sigma}_{bc}$, $\hat{\sigma}_{db}^0=\hat{\sigma}_{db}e^{i\nu_d t}$, $\hat{\sigma}_{dc}^0=\hat{\sigma}_{dc}e^{i\nu_d t}$, $\hat{\sigma}_{da}^0=\hat{\sigma}_{da}e^{i(\nu_d-\nu) t}$, $\hat{a}_j=\hat{a}_j e^{-i\nu t}$, the evolution of the quantum fields as they pass through the atomic medium is given as
\begin{equation}\label{eq3}
\begin{split}
&\frac{\partial\hat{a}_\pm}{\partial z}+\frac{1}{c}\frac{\partial\hat{a}_\pm}{\partial t}=\frac{-iNcg_c}{l}\hat{\sigma}^{(1)}_{\pm},
\end{split}
\end{equation}
where $g_c=(-\wp/\hbar)\sqrt{\hbar\nu/2\epsilon_0Ac}$, with $A$ as area of cross-section, $\hat{\sigma}_{\pm}=(\hat{\sigma}_{ca}\pm\hat{\sigma}_{ba})/\sqrt{2}$, and $\hat{a}_\pm=(\hat{a}_{2}\pm\hat{a}_{1})/\sqrt{2}$. Assuming the quantum fields are weak, they are treated perturbatively up to first order (represented by the superscript $`(1)$') in Eq. \eqref{eq3}. In this work, we further assume an undepleted drive approximation. The output quantum fields from the vapor cell are determined by the atomic operators $\hat{\sigma}_{ba}^{(1)}$ and $\hat{\sigma}_{ca}^{(1)}$. The dynamics of $\hat{\sigma}_{xy}$ can be obtained from the Heisenberg-Langevin equation \cite{gardiner2004quantum} (please see appendix . \ref{appendix}). These operator equations in Eq. \eqref{Eqap-1} to Eq. \eqref{Eqap-10} in the appendix. \ref{appendix} form a self-consistent set that can be solved to obtain the population and coherence dynamics.

\begin{figure}[hbt!]
\centering
\includegraphics[scale=1.3]{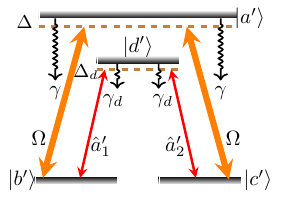}
\caption{The atomic levels are same as in Figure. \ref{fig1}. The transitions $|a\rangle-|b\rangle$ and $|a\rangle-|c\rangle$ are driven at Rabi frequency $\Omega$ by classical driving lasers, while $|d\rangle-|b\rangle$ and $|d\rangle-|c\rangle$ are coupled to weak quantum fields described by annihilation operators $\hat{a}'_1$ and $\hat{a}'_2$, respectively.}
\label{fig11}
\end{figure}

The level $|a\rangle$ is coupled to $|b\rangle$ and $|c\rangle$ through weak quantum fields. The weak fields cannot excite any population to the level $|a\rangle$ and hence $\bar{\sigma}_{aa}=0$ always, where $\bar{\sigma}_{xy}$ represents a steady state solution of the zeroth order of $\hat{\sigma}_{xy}$. The driving lasers are always on two-photon resonance, and they drive the transitions $|d\rangle-|b\rangle$ and $|d\rangle-|c\rangle$ at the same Rabi frequency $\Omega$, the population in $|b\rangle$, $|c\rangle$ and $|d\rangle$ is given as 
\begin{equation}\label{sigmabb}
\begin{split}
\bar{\sigma}_{bb}=\bar{\sigma}_{cc}=&\frac{\gamma_{bc}(\Delta_d^2+\gamma_d^2)+2|\Omega|^2(\gamma_{bc}+\gamma_d)}{2[\gamma_{bc}(\Delta_d^2+\gamma_d^2)+(3\gamma_{bc}+2\gamma_d)|\Omega|^2]},\quad \bar{\sigma}_{dd}=1-(\bar{\sigma}_{bb}+\bar{\sigma}_{cc}),
\end{split}
\end{equation}
where $2\Omega=-\wp E_0/\hbar$ is the Rabi frequency, $\Delta_d=(\omega_{d}-\omega_b)-\nu_d$, $\gamma_{bc}$ is the decoherence on the forbidden transition $|b\rangle - |c\rangle$, and $\gamma_d$ is the population decay rate from $|d\rangle$ to $|b\rangle$ and $|c\rangle$. Note that $\bar{\sigma}_{bb}=\bar{\sigma}_{cc}=1/2$ when $\gamma_{bc}=0$ as the classical driving lasers always on the two-photon resonance condition. The coherence in the $|b\rangle-|c\rangle$ transition is independent of quantum fields as two weak fields cannot create the zeroth-order two-photon coherence $\bar{\sigma}_{bc}$. Hence $\bar{\sigma}_{bc}$ is given as
\begin{equation}\label{sigmabc}
\bar{\sigma}_{bc}=\frac{-\gamma_d|\Omega|^2}{\gamma_{bc}(\Delta_d^2+\gamma_d^2)+(3\gamma_{bc}+2\gamma_d)|\Omega|^2}.
\end{equation}
Note that $\bar{\sigma}_{bc}=-1/2$ when $\gamma_{bc}=0$, as the driving lasers are always on two-photon resonance. The zeroth order coherence on $|d\rangle-|b\rangle$ and $|d\rangle-|c\rangle$ transitions is given as
\begin{equation}\label{sigmadc}
\bar{\sigma}_{dc}=\bar{\sigma}_{db}=\frac{i\gamma_{bc}(i\Delta_d+\gamma_d)\Omega^*}{2[\gamma_{bc}(\Delta_d^2+\gamma_d^2)+(3\gamma_{bc}+2\gamma_d)|\Omega|^2]}.
\end{equation}
The Eq. \eqref{sigmabb}, Eq. \eqref{sigmabc}, and Eq. \eqref{sigmadc}, together show that the population and coherence dynamics of the lambda system spanned by $|d\rangle$, $|c\rangle$ and $|b\rangle$ are completely independent of the quantum fields. However, the zeroth order coherences and populations influence the quantum coherence as
\begin{equation}\label{sigmapl}
\begin{split}
\dot{\hat{\sigma}}_+^{(1)}=\Gamma\hat{\sigma}_++ig_c\hat{a}_+\bar{n}-ig_c\hat{a}_+\bar{\sigma}_{bc}+i\sqrt{2}\Omega\hat{\sigma}_{da}^{(1)}+\frac{\hat{F}_{ba}+\hat{F}_{ca}}{\sqrt{2}},
\end{split}
\end{equation}
\begin{equation}\label{sigmami}
\dot{\hat{\sigma}}_-^{(1)}=\Gamma\hat{\sigma}_-+ig_c\hat{a}_-\bar{n}+ig_c\hat{a}_-\bar{\sigma}_{bc}+\frac{\hat{F}_{ca}-\hat{F}_{ba}}{\sqrt{2}},
\end{equation}
\begin{equation}\label{sigmada}
\dot{\hat{\sigma}}_{da}^{(1)}=\Gamma_{da}\hat{\sigma}_{da}^{(1)}-ig_c\bar{\sigma}_{db}\sqrt{2}\hat{a}_++i\Omega^*\sqrt{2}\hat{\sigma}_+^{(1)}+\hat{F}_{da},
\end{equation}
where $\bar{n}=\bar{\sigma}_{aa}-\bar{\sigma}_{bb}=\bar{\sigma}_{aa}-\bar{\sigma}_{cc}$, $\hat{F}_{xy}$ is the atomic noise operator, $\Gamma=-i\Delta-\gamma$, $\Gamma_{da}=i(\Delta_d-\Delta)-\gamma_{da}-\gamma_d-\gamma$ with $\Delta=(\omega_{a}-\omega_b)-\nu$. Cross-correlations between the atomic noise operators are given in appendix. \ref{appendix}, and $\langle\hat F_{xy}\rangle=0$. Using the Fourier transform function $\hat{O}(\omega)=\int_{-\infty}^\infty\hat{O}(t)e^{i\omega t}dt/\sqrt{2\pi}$ with $\omega$ as Fourier frequency, Eq. \eqref{sigmapl} to Eq. \eqref{sigmada} can be solved and substituted in Eq. \eqref{eq3} to obtain
\begin{equation}\label{out11}
\frac{\partial\hat{a}_{-}(\omega)}{\partial z}=d_-\hat{a}_{-}(\omega)+\frac{iNg_c}{l}\frac{\hat{F}_{ca}(z,\omega)-\hat{F}_{ba}(z,\omega)}{\sqrt{2}(i\omega+\Gamma)},
\end{equation}
\begin{equation}\label{out12}
\begin{split}
&\frac{\partial\hat{a}_{+}(\omega)}{\partial z}=d_+\hat{a}_{+}(\omega)+\frac{iNg_c}{l}\frac{(i\omega+\Gamma_{da})(\hat{F}_{ba}(z,\omega)+\hat{F}_{ca}(z,\omega))-i2\Omega\hat{F}_{da}(z,\omega)}{\sqrt{2}[(i\omega+\Gamma)(i\omega+\Gamma_{da})+2|\Omega|^2]},
\end{split}
\end{equation}
where
\begin{equation*}
\begin{split}
&d_-=\frac{i\omega}{c}+\frac{\kappa(-\bar{\sigma}_{bc}-\bar{n})}{i\omega+\Gamma},\quad d_+=\frac{i\omega}{c}+\frac{\kappa[(\bar{\sigma}_{bc}-\bar{n})(i\omega+\Gamma_{da})-i2\Omega\bar{\sigma}_{db}]}{(i\omega+\Gamma)(i\omega+\Gamma_{da})+2|\Omega|^2},
\end{split}
\end{equation*}
$\kappa=Ng_c^2/l$. Now, consider the second system shown in Figure. \ref{fig11}. The only difference between Figure. \ref{fig1} and Figure. \ref{fig11} is that the $|a'\rangle-|b'\rangle$ ($|a'\rangle-|c'\rangle$) transition is driven at Rabi frequency $\Omega$ by driving lasers, while the $|d'\rangle-|b'\rangle$ ($|d'\rangle-|c'\rangle$) transition is coupled to a weak quantum field described by annihilation operator $\hat{a}'_1$ ($\hat{a}'_2$). We added a ‘prime’ in the superscript to differentiate the atom-laser interaction in Figure. \ref{fig1} from the atom-laser interaction in Figure. \ref{fig11}. The atomic-level structure in Figure. \ref{fig11} is the same as in Figure. \ref{fig1}. Hence, we can adopt the Eq. \eqref{out11} and Eq. \eqref{out12} for Figure. \ref{fig11} and write the evolution of operators $\hat{a}'_1$ and $\hat{a}'_2$ as 
\begin{equation}\label{out11-prime}
\frac{\partial\hat{a}'_{-}(\omega)}{\partial z}=d'_-\hat{a}'_{-}(\omega)+\frac{iNg_c}{l}\frac{\hat{F}'_{ca}(z,\omega)-\hat{F}'_{ba}(z,\omega)}{\sqrt{2}(i\omega+\Gamma'_d)},
\end{equation}
\begin{equation}\label{out12-prime}
\begin{split}
&\frac{\partial\hat{a}'_{+}(\omega)}{\partial z}=d'_+\hat{a}'_{+}(\omega)+\frac{iNg_c}{l}\frac{(i\omega+\Gamma'_{da})(\hat{F}'_{ba}(z,\omega)+\hat{F}'_{ca}(z,\omega))-i2\Omega\hat{F}'_{da}(z,\omega)}{\sqrt{2}[(i\omega+\Gamma'_d)(i\omega+\Gamma'_{da})+2|\Omega|^2]},
\end{split}
\end{equation}
where
\begin{equation*}
\begin{split}
&d'_-=\frac{i\omega}{c}+\frac{\kappa(-\bar{\sigma}'_{bc}-\bar{n}')}{i\omega+\Gamma'}, \quad d'_+=\frac{i\omega}{c}+\frac{\kappa[(\bar{\sigma}'_{bc}-\bar{n}')(i\omega+\Gamma'_{da})-i2\Omega\bar{\sigma}'_{db}]}{(i\omega+\Gamma')(i\omega+\Gamma'_{da})+2|\Omega|^2},
\end{split}
\end{equation*}
\begin{equation*}
\begin{split}
&\bar{\sigma}'_{bc}=\frac{-\gamma|\Omega|^2}{\gamma_{bc}(\gamma{^2}+\Delta^2)+(3\gamma_{bc}+2\gamma)|\Omega|^2},\quad \bar{n}'=-\frac{\gamma_{bc}(\gamma{^2}+\Delta^2)+2(\gamma_{bc}+\gamma)|\Omega|^2}{2\gamma_{bc}(\gamma{^2}+\Delta^2)+2(3\gamma_{bc}+2\gamma)|\Omega|^2},
\end{split}
\end{equation*}
\begin{equation*}
\bar{\sigma}'_{db}=\bar{\sigma}'_{dc}=\frac{i\gamma_{bc}(i\Delta+\gamma)\Omega^*}{2[\gamma_{bc}(\Delta^2+\gamma^2)+(3\gamma_{bc}+2\gamma)|\Omega|^2]}.
\end{equation*}
where $\Gamma'=-i\Delta_d-\gamma_d$, $\Gamma'_{da}=i(\Delta-\Delta_d)-\gamma_{da}-\gamma-\gamma_d$, $\hat{a}'_{\pm}=(\hat{a}'_2\pm\hat{a}'_1)/\sqrt{2}$. All the primed variables have the same meaning as the unprimed variables but for the atom-laser interaction shown in Figure. \ref{fig11}. We further assumed that the sample length $l$, atomic number density, and coupling constant $g_c$ are same in Figure. \ref{fig1} and Figure. \ref{fig11}.
\section{Clock design}\label{clock-design}
\begin{figure*}[bth!]
\centering
\includegraphics[width=0.9\linewidth]{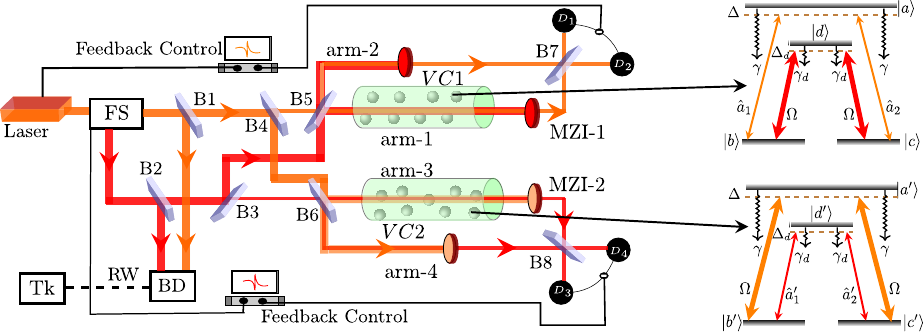}
\caption{Schematics of the optical atomic clock. Vapor cell VC1 (VC2) containing Rb atoms simulating atom-laser interaction as shown in Figure. \ref{fig1} in MZI-1 (Figure. \ref{fig11} in MZI-2), orange/red ellipse: driving laser frequency filters, FS : frequency shifter, Bi (i=1,2,...,8): beam splitters, $D_i$ ($i=1,2,3,4$): photo-detectors, RW: radio-wave signal, BD: beat detector, Tk: Timekeeping.}
\label{fig2}
\end{figure*}
The schematic of the clock is shown in Figure. \ref{fig2}. A laser field with frequency $\nu$ goes through a frequency shifter (FS). The FS can be acousto/electro optic modulator or electro-optic frequency shifter. The output from the FS contains an unmodulated laser field (shown by thick orange line) at frequency $\nu$ and a modulated laser field (shown by thick red line) at frequency $\nu_d$. A small part of the unmodulated laser field is reflected from beam splitter B1 and mixed with a small part of the modulated laser field reflected from beam splitter B2. A beat signal at frequency $\nu-\nu_d$ is measured at beat detector (BD) by mixing the reflected lasers from B1 and B2. Then, the laser field at frequency $\nu \;(\nu_d)$ is split into two parts using a high reflective beam splitter B4 (B3). The transmitted field from B4 and reflected field from B3 are sent into the Mach-Zehnder interferometer (MZI-1) formed by arm-1 and arm-2. Similarly, the reflected field from B4 and transmitted field from B3 are sent into another Mach-Zehnder interferometer (MZI-2) formed by arm-3 and arm-4. The reflected fields from the B4 and B3 have equal electric field amplitude $2E_0$ and are treated classically. We refer to these fields as driving lasers. The annihilation
operators for the transmitted fields from B4 and B3 are represented by $\hat{E}$ and $\hat{E}'$, respectively. The quantum fields are weak, $i.e.,$ $\langle\hat{E}^\dagger\hat{E}\rangle=\langle\hat{E}'^\dagger\hat{E}'\rangle\ll 4|E_0|^2$. Note that the frequencies of the weak quantum fields and driving lasers are $\nu$ and $\nu_d$, respectively, in MZI-1. On the other hand, the frequencies of the weak quantum fields and driving lasers are $\nu_d$ and $\nu$, respectively, in MZI-2.
\subsection{Signal and noise of the MZI-1}\label{SMZI-1}
The atomic vapor and the laser fields in the MZI-1 vapor cell (VC1) are tailored to simulate the atom-laser interaction, as in Figure. \ref{fig1}. The annihilation operator for the quantum field entering the VC1 is $(\hat{E}+i\hat{V})/\sqrt{2}$, where $\hat{V}$ is the annihilation operator for the vacuum entering MZI-1. The linearly polarized quantum field entering the VC1 is split into a left circular polarized field, represented with annihilation operator $\hat{a}_{10}$, and a right circular polarized field, represented with annihilation operator $\hat{a}_{20}$, respectively. The polarization and frequency $\nu$ are chosen such that $\hat{a}_{10}$ and $\hat{a}_{20}$ couple $|a\rangle-|b\rangle$ and $|a\rangle-|c\rangle$ transitions, respectively. Similarly, the linearly polarized classical driving laser at the input of the VC1 is split into a left circular field and a right circular field with equal amplitude. The $|d\rangle-|b\rangle$ and $|d\rangle-|c\rangle$ transitions are driven 
 by the left circular driving laser and right circular driving laser, respectively, at same Rabi frequency $2\Omega$. We assume the quantum fields are weak enough and treat them perturbatively up to first order. Hence, the propagation of quantum fields in VC1 is described by Eq. \eqref{out11} and Eq. \eqref{out12}. Solving Eq. \eqref{out11} and Eq. \eqref{out12} gives 
 \begin{equation}\label{out15-1}
\begin{split}
&\hat{a}_{+l}(\omega)=e^{ld_+}\hat{a}_{+0}(\omega)+\frac{iNg_c}{l}\int\limits_{-l/2}^{l/2}dz_1 e^{(l/2-z_1)d_+}\frac{(i\omega+\Gamma_{da})[\hat{F}_{ba}(z_1,\omega)+\hat{F}_{ca}(z_1,\omega)]-i2\Omega\hat{F}_{da}(z_1,\omega)}{\sqrt{2}[(i\omega+\Gamma)(i\omega+\Gamma_{da})+2|\Omega|^2]},
\end{split}
\end{equation}
\begin{equation}\label{out16-1}
\begin{split}
\hat{a}_{-l}(\omega)&=e^{ld_-}\hat{a}_{-0}(\omega)+\frac{iNg_c}{l}\int\limits_{-l/2}^{l/2}dz_1 e^{(l/2-z_1)d_-}\frac{\hat{F}_{ca}(z_1,\omega)-\hat{F}_{ba}(z_1,\omega)}{\sqrt{2}(i\omega+\Gamma)},
\end{split}
\end{equation}
where $\hat{a}_{\pm l}(\omega)=[\hat{a}_{2l}(\omega)\pm\hat{a}_{1l}(\omega)]/\sqrt{2}$, $\hat{a}_{jl}=\hat{a}_j(l,\omega)$, $\hat{a}_{j0}=\hat{a}_j(0,\omega)$. When $|\Omega|^2=0$, $d_+=d_-=-\kappa/[2(i\Delta+\gamma)]$, and the quantum fields get absorbed by the atomic medium at $\Delta=0$. Hence, the driving lasers are necessary to suppress the absorption of the quantum fields through Electromagnetically induced transparency \cite{PhysRevLett.66.2593}.
The driving lasers are eliminated at the
output of the VC1 using a frequency filter. The annihilation operator for the quantum field reaching B7 from arm-1 is $(\hat{a}_{+l}+i\hat{a}_{-l})/\sqrt{2}$. As the goal of the clock is to correct the smallest measurable $\Delta$ and $\Delta_d$ values, $\Delta\ll\gamma$ and $\Delta_d\ll\gamma$ are the regimes of interest in this work. For $\Delta_d\ll\gamma$, we approximate $\bar{\sigma}_{bc}+\bar{n}\approx 1$ and $ld_-\approx-l\kappa/\Gamma$ at $\omega=0$. Hence the real part $\mathcal{R}(ld_-)$ of $ld_-$, gives a Lorentzian dip with minimum value of $-l\kappa/\gamma$ at $\Delta=0$, and half-minima at $\Delta=\pm\gamma$. Assuming $l=0.11$ m, $\gamma/2\pi=6.06$ MHz and number density of atoms $2\times 10^{18}$ m$^{-3}$ gives $-l\kappa/\gamma\approx -16265.4$. Hence $e^{ld_-}\rightarrow 0$ for $\Delta\ll\gamma$ and $\Delta_d\ll\gamma$.

\begin{figure}[hbt!]
    \centering
    \includegraphics[width=0.5\columnwidth]{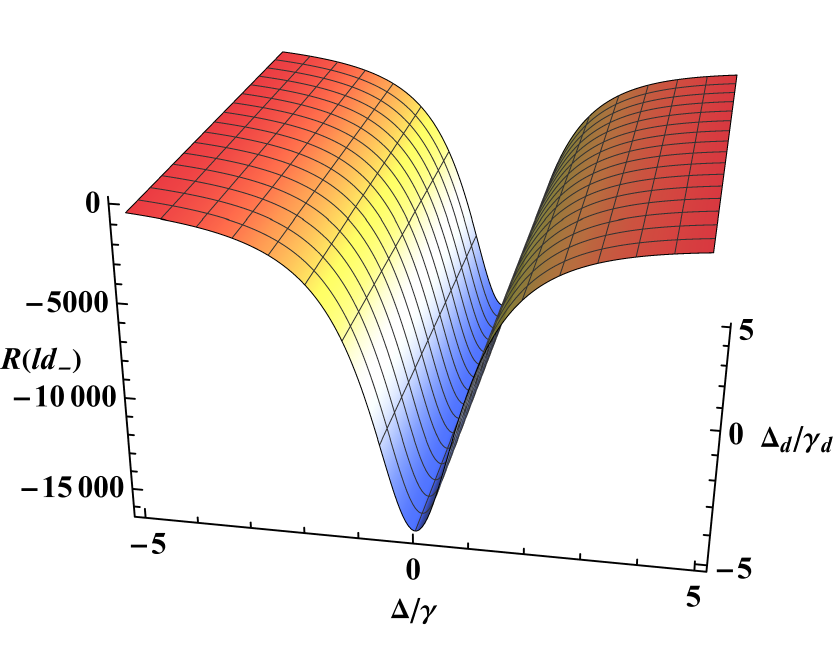}
    \caption{Variation of real part of $ld_-$ as a function of $\Delta$ and $\Delta_d$. Simulation parameters:  $\gamma_d/2\pi=6.06$ MHz, $\gamma=\gamma_d$, $|\Omega|=2\gamma$ (corresponding to $54$ mW optical power), $\gamma_{bc}=10^{-3}\gamma$, $\gamma_{da}=\gamma_{bc}$, $l=0.11$ m, atomic number density $2\times 10^{18}$ m$^{-3}$, $\omega=0$.}
    \label{d_1l0_real}
\end{figure}

For more clarity, $\mathcal{R}(ld_-)$, is plotted in Figure. \ref{d_1l0_real} as a function of $\Delta$ and $\Delta_d$ considering realistic parameters ( $\gamma=\gamma_d$, $\gamma_{bc}=\gamma_{da}=10^{-3}\gamma$, $\Omega=2\gamma$, $\kappa=5.62\times 10^{12}$ Hz/m) of the $\ce{^{87}_{}Rb}$ D2 line.
Figure. \ref{d_1l0_real} shows that $|\mathcal{R}(ld_{-})|\gg 1$ and $\mathcal{R}(ld_{-})$ is negative for $\Delta\ll\gamma$ and $\Delta_d\ll\gamma$. Thus $\langle \hat{a}_{-l}\rangle\rightarrow 0$ and the average field reaching the B7 is approximated to $e^{ld_+}\langle \hat{a}_{+0}\rangle $, for $\Delta\ll\gamma$ and $\Delta_d\ll\gamma$. At this point, it will be useful to re-write $d_+$ as $d_+=d_{+r}+i(\Delta d_{+1}+\Delta_d d_{+2})$ by assuming $\Delta\ll\gamma$ and $\Delta_d\ll\gamma$, where
\begin{equation*}
\begin{split}
&d_{+r}=\frac{\kappa\gamma_{bc}[-\gamma_d^2\gamma_o-2|\Omega|^2(\gamma+\gamma_{da})]}{2[\gamma_{bc}(\gamma_d^2+3|\Omega|^2)+2|\Omega|^2\gamma_d][\gamma\gamma_o+2|\Omega|^2]},\\
& d_{+1}=\kappa\gamma_{bc}\frac{{[\gamma_d^2 \gamma_o^2- 4 |\Omega|^4}{ + 
 2 (\gamma^2 - \gamma_d^2 + 
    2 \gamma \gamma_{da} + \gamma_{da} [\gamma_d + \gamma_{da}]) \
|\Omega|^2 ]}}{2[\gamma_{bc} (\gamma_d^2 + 3 |\Omega|^2) + 
      2 |\Omega|^2\gamma_d ] [\gamma \gamma_o + 
   2 |\Omega|^2]^2 },\\
&d_{+2}=\frac{ \kappa \gamma_{bc} |\Omega|^2 [(\gamma + \gamma_d)^2 + \gamma \gamma_{da} + 4 |\Omega|^2]}{ [\gamma_{bc}( \gamma_d^2 + 3|\Omega|^2) + 
      2 |\Omega|^2\gamma_d][\gamma \gamma_o + 
   2 |\Omega|^2]^2},
\end{split}
\end{equation*}
$\gamma_o=\gamma+\gamma_d+\gamma_{da}$. The annihilation operator for the field reaching the B7 from arm-2 is given as $(\hat{E}-i\hat{V})/\sqrt{2}$ after adding a $-\pi/2$ phase to it. The output at the MZI-1 is obtained by subtracting the intensity measured at the photo-detectors $D_1$ and $D_2$. The average signal $\mathcal{S}$ and the quantum noise $\mathcal{N}$ in the MZI-1 output are given as
\begin{equation}\label{Sig-Noise-1}
\begin{split}
&\mathcal{S}=\frac{e^{ld_{+r}}}{2}|\bar{E}|^2l\left(d_{+1}\Delta+d_{+2}\Delta_d\right),\quad \mathcal{N}=\sqrt{\frac{|\bar{E}|^2}{2}\left(1+\frac{e^{2ld_{+r}}}{4}\right)},
\end{split}
\end{equation}
where $|\bar{E}|^2=\langle\hat{E}^\dagger\hat{E}\rangle$.
\subsection{Signal and noise of the MZI-2}\label{sec-interferometer}
The atomic vapor and laser fields in the MZI-2 vapor cell (VC2) are chosen such that the atom-laser interaction shown
in Figure. \ref{fig11}. The quantum field entering the vapor cell in arm-3 is described as $(\hat{E}'+i\hat{V}')/\sqrt{2}$, where $\hat{V}'$ is the vacuum annihilation operator entering MZI-2. The linearly polarized quantum field entering the vapor cell is split into a left circular polarized field, represented by annihilation operator $\hat{a}'_{10}$, and a right circular polarized field, represented with annihilation operator $\hat{a}'_{20}$, respectively. The polarization and frequency $\nu_d$ are chosen such that $\hat{a}'_{10}$ and $\hat{a}'_{20}$ couple $|d'\rangle-|b'\rangle$ and $|d'\rangle-|c'\rangle$ transitions, respectively. Similarly, the linearly polarized classical driving laser at the input of the vapor cell is split into a left circular field and a right circular field with equal amplitude. The $|a'\rangle-|b'\rangle$ and $|a'\rangle-|c'\rangle$ transitions are driven at Rabi frequency $2\Omega$ by the left circular driving laser and right circular driving laser, respectively. We assume the quantum fields are weak enough and treat them perturbatively up to first order. Hence, the propagation of quantum fields in VC2 is described by Eq. \eqref{out11-prime} and Eq. \eqref{out12-prime}. Solving Eq. \eqref{out11-prime} and Eq. \eqref{out12-prime} gives
\begin{equation}\label{out15}
\begin{split}
&\hat{a}'_{+l}(\omega)=e^{ld'_+}\hat{a}'_{+0}(\omega)+\frac{iNg_c}{l}\int\limits_{-l/2}^{l/2}dz_1 e^{(l/2-z_1)d'_+}\frac{(i\omega+\Gamma'_{da})[\hat{F}'_{ba}(z_1,\omega)+\hat{F}'_{ca}(z_1,\omega)]-i2\Omega\hat{F}'_{da}(z_1,\omega)}{\sqrt{2}[(i\omega+\Gamma')(i\omega+\Gamma'_{da})+2|\Omega|^2]},
\end{split}
\end{equation}
\begin{equation}\label{out16}
\begin{split}
&\hat{a}'_{-l}(\omega)=e^{ld'_-}\hat{a}'_{-0}(\omega)+\frac{iNg_c}{l}\int\limits_{-l/2}^{l/2}dz_1 e^{(l/2-z_1)d'_-}\frac{\hat{F}'_{ca}(z_1,\omega)-\hat{F}'_{ba}(z_1,\omega)}{\sqrt{2}(i\omega+\Gamma')},
\end{split}
\end{equation}
where $\hat{a}'_{\pm l}(\omega)=[\hat{a}'_{2l}(\omega)\pm\hat{a}'_{1l}(\omega)]/\sqrt{2}$, $\hat{a}'_{jl}=\hat{a}'_j(l,\omega)$, $\hat{a}'_{j0}=\hat{a}'_j(0,\omega)$. Again, the driving lasers are necessary to suppress the absorption of the quantum fields, are eliminated at the output of the VC2 using a frequency
filter. Plot for the $\mathcal{R}(ld'_-)$ is same as $\mathcal{R}(ld_-)$ as the atomic medium in both the vapor cells are same. Hence, just like in MZI-1, $\langle\hat{a}'_{-l}\rangle\rightarrow 0$ and only $\hat{a}'_{+l}$ contributes to the signal at the MZI-2 output. We re-write $d'_+$ as $d'_+=d'_{+r}+i\left(\Delta_d d'_{+r}+\Delta d'_{+2}\right)$ by assuming $\Delta\ll\gamma$ and $\Delta_d\ll\gamma$. The signal $\mathcal{S}'$ and the quantum noise $\mathcal{N}'$ in MZI-2 output are given as
\begin{equation}\label{sens-second}
\begin{split}
&\mathcal{S}'=\frac{e^{ld'_{+r}}}{2}|\bar{E}'|^2l\left(d'_{+1}\Delta_d+d'_{+2}\Delta\right),\quad \mathcal{N}'=\sqrt{\frac{|\bar{E}'|^2}{2}\left(1+\frac{e^{2ld'_{+r}}}{4}\right)},
\end{split}
\end{equation}
where $|\bar{E'}|^2=\langle\hat{E'}^\dagger\hat{E'}\rangle$,
\begin{equation*}
\begin{split}
&d'_{+r}=\frac{\kappa\gamma_{bc}[-\gamma^2\gamma_o-2|\Omega|^2(\gamma_d+\gamma_{da})]}{2[\gamma_{bc}(\gamma^2+3|\Omega|^2)+2|\Omega|^2\gamma][\gamma_d\gamma_o+2|\Omega|^2]},\\
& d'_{+1}=\kappa\gamma_{bc}\frac{{[\gamma^2 \gamma_o^2- 4 |\Omega|^4}{ + 
 2|\Omega|^2 (\gamma_d^2 - \gamma^2 + 
    2 \gamma_d \gamma_{da} + \gamma_{da} [\gamma + \gamma_{da}]) \
 ]}}{2[\gamma_{bc} (\gamma^2 + 3 |\Omega|^2) + 
      2 |\Omega|^2\gamma] [\gamma_d \gamma_o + 
   2 |\Omega|^2]^2 },\\
&d'_{+2}=\frac{ \kappa \gamma_{bc} |\Omega|^2 [(\gamma_d + \gamma)^2 + \gamma_d \gamma_{da} + 4 |\Omega|^2]}{ [\gamma_{bc}( \gamma^2 + 3|\Omega|^2) + 
      2 |\Omega|^2\gamma][\gamma_d \gamma_o + 
   2 |\Omega|^2]^2}.
\end{split}
\end{equation*}
\subsection{Setting $\Delta=\Delta_d\approx0$, stability of the clock, correcting FS}\label{eofs}
\begin{figure}[hbt!]
    \centering
\subfloat[\label{d_2l0_real}\centering ]{\includegraphics[ width=0.5\columnwidth]{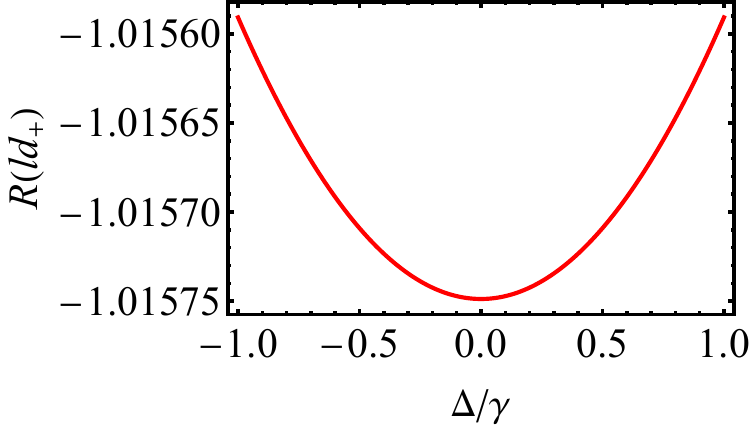} }
     \hfill
    \subfloat[\label{im_d2l}\centering]{\includegraphics[width=0.45\columnwidth]{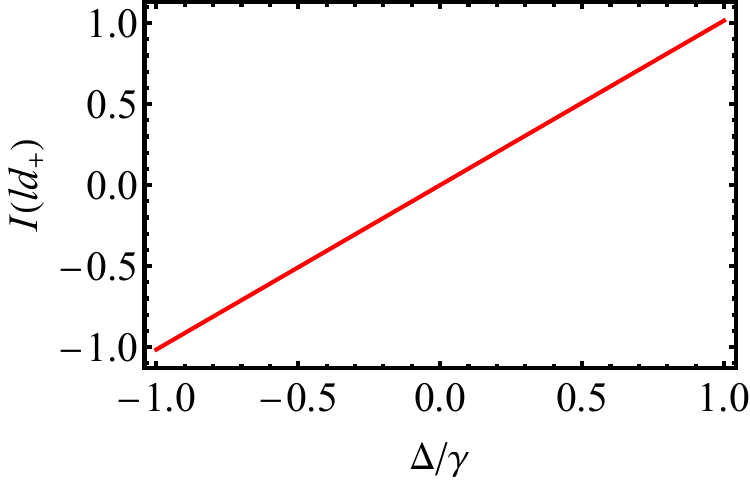}}\par
\caption{Variation of real part of $ld_+$ (\ref{d_2l0_real}), and imaginary part of $ld_+$ (\ref{im_d2l}) as a function of $\Delta$ for $\Delta=\Delta_d$. All other simulation parameters are same as Figure. \ref{d_1l0_real}.}%
    \label{d_1l0}%
\end{figure}

The $\mathcal{S}$ in Eq. \eqref{Sig-Noise-1} and $\mathcal{S}'$ in Eq. \eqref{sens-second} are a function of $\Delta$ and $\Delta_d$. The reason for setting up two interferometers in Figure. \ref{clock-design} is that we need two equations to solve for two unknown variables: $\Delta$ and $\Delta_d$. The $\mathcal{S}$ and $\mathcal{S}'$ provide such required equations. For the D2 line of alkali metals (Cs, Rb, Na,...), $\gamma\approx\gamma_d$, hence $d_{+r}=d'_{+r}$ and $d_{+j}=d'_{+j}$. By taking $|\bar{E}|^2=|\bar{E}'|^2$, we can have $\mathcal{S}=\mathcal{S}' \iff \Delta=\Delta_d$. Hence forcing both the interferometers to have same signal ensures that $\Delta=\Delta_d$ condition is imposed. For $\Delta=\Delta_d\lessapprox\gamma$, the variation of real part $\mathcal{R}(ld_+)$ and imaginary part $\mathcal{I}(ld_+)$ of $ld_+$ is shown in Figure. \ref{d_2l0_real} and Figure. \ref{im_d2l}, respectively (note that the plots of $\mathcal{R}(ld'_+)$ and $\mathcal{I}(ld'_+)$ are the same as the plots of $\mathcal{R}(ld_+)$ and $\mathcal{I}(ld_+)$, respectively, hence same conclusion apply to MZI-2 as well). Figure. \ref{im_d2l} shows that the $\mathcal{I}(ld_+)$ varies linearly around $\Delta\approx0$.
Whenever the laser frequencies $\nu$ and $\nu_d$ are detuned from the atomic resonance frequencies $\omega_{ab}$ and $\omega_{db}$ by same amount (i.e., $\Delta=\Delta_d\ll\gamma$), a non-zero phase is added to $\langle \hat{a}_{+l}\rangle$. This additional phase shifts the interference fringes, which is used as feedback to set $\Delta=\Delta_d\approx0$.
Using Eq. \eqref{Sig-Noise-1} and Eq. \eqref{sens-second}, the total noise is calculated as $\sqrt{\mathcal{N}^2+\mathcal{N'}^2}$. Then the stability $\Delta_s/\nu_d$, where $\Delta_s$ as the frequency sensitivity, of the clock is given as
\begin{equation}\label{delta-sen-3}
\frac{\Delta_s}{\nu_d}=\frac{\sqrt{1+4e^{-2ld_{+r}}}}{\nu_d l\left(d_{+1}+d_{+2}\right)\sqrt{|\bar{E}|^2}}.
\end{equation}
For the considered atomic levels, described in Figure. \ref{fig1}, $\nu_d/2\pi=3.84\times 10^{14}$ Hz. Having theoretically described how to use dual interferometers signals to set $\Delta=\Delta_d\; (\mathcal{S}=\mathcal{S}'\iff\Delta=\Delta_d$), we now describe how the $\Delta=\Delta_d$ condition could be achieved in the experiment. The modulated laser field frequency is related \cite{Drain1972,Zeng:20} to the unmodulated laser frequency as $\nu_d=\nu-\nu_0$, where $\nu_0$ is the modulating frequency of the FS. Lets assume that the frequency $\nu$ ($\nu_d$) is tuned closed to $\omega_{ab}$ $(\omega_{db})$ such that $\Delta\ll\gamma$ ($\Delta_d\ll\gamma$). For small $\Delta$ and $\Delta_d$, the signal from the MZI-1 is equal to the signal from MZI-2 if and only if $\Delta=\Delta_d$. Hence, whenever $\mathcal{S}\neq \mathcal{S}'$, a feedback loop corrects the
FS modulating frequency $\nu_0$ until $\mathcal{S}=\mathcal{S}'$. This enforces the condition $\Delta=\Delta_d$ in the experiment. After calibrating the FS, another
feedback loop from the MZI-1 will adjust the laser frequency such that $\Delta=\Delta_d\approx0$. This adjusts both $\nu$ and $\nu_d$ to $\omega_{ab}$ and $\omega_{db}$ resonances, respectively.
\subsection{Simplified picture}\label{simp-pic}
The expressions in Eq. \eqref{Sig-Noise-1} and Eq. \eqref{sens-second} are intractably large once $d_{\pm}$ and $d'_{\pm}$ are substituted in it. In order to obtain an intuitive understanding of the results described in subsection \ref{SMZI-1}, \ref{sec-interferometer} and \ref{eofs}, we resort to approximation by taking $|\Omega|^2\gg\gamma^2\approx\gamma_d^2$. Such approximation is not required for the operation of the clock, but they will simplify  Eq. \eqref{Sig-Noise-1}, Eq. \eqref{sens-second} and Eq. \eqref{delta-sen-3} to understand the clock’s operation. Assuming $|\Omega|^2\gg\gamma^2$, we write $ld_-\approx -(\kappa l/\gamma)(1-i\Delta/\gamma)$, and $ld_+\approx -(\kappa\gamma_{bc} l/4|\Omega|^2)[((\gamma+\gamma_{da})/\gamma)+i((\Delta-2\Delta_d)/2\gamma)]$ for small $\Delta$ and $\Delta_d$. By adjusting the driving laser intensity such that $4|\Omega|^2\approx\kappa\gamma_{bc}l$, we ensure that $|\mathcal{R}(ld_-)|\approx4|\Omega|^2/\gamma\gamma_{bc}\ggg 1$ and $\mathcal{R}(ld_-)<0$. Hence, $e^{ld_{-}}\rightarrow 0$, and we can write
\begin{equation}\label{new-1}
\begin{split}
    &\bar{a}_{-l}=\bar{a}_{-0}\frac{e^{ld_-}}{2}\approx 0,\quad\bar{a}_{+l}=\bar{a}_{+0}\frac{e^{ld_+}}{2}\approx \frac{\bar{a}_{+0}}{2}e^{-\frac{\kappa\gamma_{bc}l}{2|\Omega|^2}\left(\frac{\gamma+\gamma_{da}}{\gamma_d}+i\frac{\Delta-2\Delta_d}{2\gamma_d}\right)}.
    \end{split}
\end{equation}
Hence, only $\bar{a}_{+l}$ contributes to the average signal at MZI-1 output. The Eq. \eqref{new-1} implies that the $\mathcal{I}(ld_+)$ and  $\mathcal{R}(ld_+)$ represents the phase $\phi$ and absorption, respectively, of $\hat{a}_{+l}$.

In MZI-2, again by assuming $|\Omega|^2\gg\gamma^2\approx\gamma_d^2$, we approximate $ld'_-\approx -(\kappa l/\gamma_d)(1-i\Delta_d/\gamma_d)$, and $ld'_+\approx -(\kappa\gamma_{bc} l/4|\Omega|^2)[(\gamma_d+\gamma_{da})/\gamma_d+i(\Delta_d-2\Delta)/2\gamma_d]$. Adjusting $4|\Omega|^2\approx\kappa\gamma_{bc}l$, ensures that $|\mathcal{R}(ld'_-)|\approx 4|\Omega|^2/\gamma_d\gamma_{bc}\ggg 1$ and $\mathcal{R}(ld'_-)<0$. Hence we can approximate $e^{ld'_-}\rightarrow 0$ and write
\begin{equation}\label{new-2}
\begin{split}
   & \bar{a}'_{-l}=\bar{a}'_{-0}e^{ld'_-}\approx 0,\quad \bar{a}'_{+l}=\bar{a}'_{+0}\frac{e^{ld'_+}}{2}\approx \frac{\bar{a}'_{+0}}{2}e^{-\frac{\kappa\gamma_{bc}l}{2|\Omega|^2}\left(\frac{\gamma_d+\gamma_{da}}{\gamma}+i\frac{\Delta_d-2\Delta}{2\gamma}\right)}.
    \end{split}
\end{equation}
The Eq. \eqref{new-2} implies that $\mathcal{I}(ld'_+)$ and $\mathcal{R}(ld'_+)$ represents the phase $\phi'$, and absorption, respectively, of $\hat{a}'_{+l}$. The average signal in MZI-1 and MZI-2 are equal when $\phi=\phi'$. By equating the imaginary parts in the exponential
of Eq. \eqref{new-1} and Eq. \eqref{new-2}, we can conclude $\phi=\phi'\iff \Delta=\Delta_d$. Hence, the condition $\Delta=\Delta_d$ can be experimentally realized by forcing the signals from both the interferometers to be equal. Hence, the stability the clock using Eq. \eqref{delta-sen-3} is approximated as
\begin{equation}\label{delta-sen-4}
  \begin{split}
\frac{\Delta_s}{\nu_d}\approx\frac{4|\Omega|^2\gamma}{\nu_d\kappa\gamma_{bc}l}\sqrt{\frac{1+4e^{\frac{\kappa\gamma_{bc}l}{2|\Omega|^2}}}{|\bar{E}|^2}}.
  \end{split}  
\end{equation}
By using the condition $4|\Omega|^2=\kappa\gamma_{bc}l$ in Eq. \eqref{delta-sen-4}, we can write the optimum stability $\Delta_{opt}/\nu_d$, (where $\Delta_{opt}$ is $\Delta_s$ at ${4|\Omega|^2=\kappa\gamma_{bc}l}$), of the clock as 
\begin{equation}
   \frac{ \Delta_{opt}}{\nu_d}=\frac{\gamma}{\nu_d}\sqrt{\frac{1+4e^{2}}{|\bar{E}|^2}}.
\end{equation} 
\subsection{Elimination of OFC}
One second is defined by counting the number of electric field oscillations of a laser field with known frequency. The oscillations of the optical laser field are so fast that current electronics can not count them directly. Hence, the laser field frequency is down-converted to micro-wave or radio-wave frequencies by beating it with a known frequency teeth of a OFC. Generation and calibration of OFC requires sophisticated setup \cite{PhysRevLett.94.193201} including a femtosecond laser source and a self-referencing setup. The optical clock proposed in this work can simultaneously calibrate $\nu$ to $\omega_{ab}$ and $\nu_d$ to $\omega_{db}$. Hence, mixing the laser fields at $\nu$ and $\nu_d$ frequencies gives a beat signal (detected at BD in Figure. \ref{fig2}) at frequency $\nu-\nu_d=\omega_{ad}$, which is in the radio frequency regime. This down-converted signal (from BD in Figure. \ref{fig2}) is used to count the oscillations to define a second without the need of a OFC. Elimination of OFC improves portability and is also less demanding on power.
\subsection{Stark effect and standard quantum limit}
\begin{figure}[hbt!]
\centering
\includegraphics[width=0.5\columnwidth]{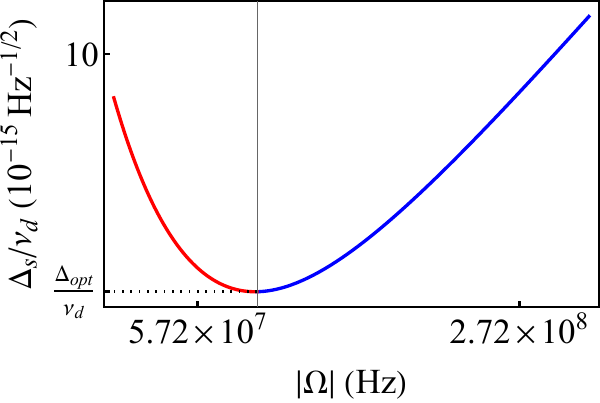}
\caption{Variation of OAC stability as a function of $|\Omega|$. The lowest point of the curve represents the optimum stability ($1.3\times 10^{-15}$ Hz$^{-1/2}$), which represents standard quantum limit, at Rabi frequency $4|\Omega|^2=\kappa\gamma_{bc}l$. Intensity of the quantum field entering the vapor cell is $|\bar{E}|^2/2=2.13\times 10^{15}$  Hz corresponding to $0.54$ mW optical power, $\nu_d/2\pi=3.84\times 10^{14}$ Hz, while all the other simulation parameters are same as Figure. \ref{d_1l0_real}.}
\label{fig3}
\end{figure}

Figure \ref{fig3} is plotted using Eq. \eqref{delta-sen-3} to show the variation of clock stability as a function of driving laser Rabi frequency $\Omega$. 
As Eq. \eqref{delta-sen-3} is too complicated, we again resort to the simplified picture discussed in the section. \ref{simp-pic}. In the simplified
picture, the stability is given by Eq. \eqref{delta-sen-4}. If $|\Omega|^2$ is decreased such that the $\kappa\gamma_{bc}l/4|\Omega|^2$ value in Eq. \eqref{delta-sen-4} increases, then the exponential term increases while the value of the term before square root decreases. As a result, the clock
stability decreases, as shown by the red part of the curve in Figure. \ref{fig3}. If $|\Omega|^2$ increases such that $\kappa\gamma_{bc}l/4|\Omega|^2$ value decreases, then the exponential term goes to 1 while the value of the term before the square root increases. Thus, again the clock
stability decreases with the increase of $\Omega$ as shown by the blue part of the curve in Figure. \ref{fig3}. Hence, the optimum stability (standard quantum limited) is achieved when the driving laser amplitude satisfies the condition $4|\Omega|^2=\kappa\gamma_{bc}l$. Qualitatively, the increase in the Rabi frequency increases the stark splitting, leading to a decrease in the dispersion and stability of the clock. If the Rabi frequency is too small, then the atomic noise (exponential term) increases leading to a reduced stability of the clock.
An important aspect of Eq. \eqref{delta-sen-3} is the $1/\sqrt{|\bar{E}|^2}$ factor, which is shot noise. Conventional optical clocks use two detection schemes: (1) Ramsey fringes \cite{PhysRevLett.79.3865}
and (2) Fluorescence measurement from two-photon excitation \cite{PhysRevApplied.12.054063}. The sensitivity of both schemes scales as $1/\sqrt{M}$ with $M$
 the number of atoms \cite{Schulte2020} and fluorescence photons \cite{PhysRevApplied.12.054063} for Ramsey fringes and fluorescence schemes, respectively. The
maximum value of $M$ is in the order of $10^6$ \cite{Farkas2010} in both schemes. The typical value of $|\bar{E}|^2$ in Eq. \eqref{delta-sen-3} can go above $10^{15}$ for a laser power of $10^{-3}$ Watt, which is significantly larger than $M$. Eventhough the Eq. \eqref{delta-sen-3} has the advantage of large $|\bar{E}|^2$, the large value of the dipole allowed transition decay rate $\gamma$ offsets the
advantage coming from $|\bar{E}|^2$. Hence, we could build a clock with a stability of approximately an order of magnitude better than the existing vapor cell clocks, which use a dipole-forbidden transition. One may explore optical non-classical states \cite{Gopinath2024} to improve the stability of the clock \cite{RevModPhys.90.035005}. For example: Using optical $N00N$ states will lead to Heisenberg scaling \cite{PhysRevA.81.030302} in Eq. \eqref{delta-sen-3}.
\section{Broadening effects}\label{broadening}
\subsection{Doppler broadening effect}\label{doppler}
\begin{figure}[hbt!]
    \centering
\subfloat[\label{d_2_real}\centering ]{\includegraphics[width=0.5\columnwidth]{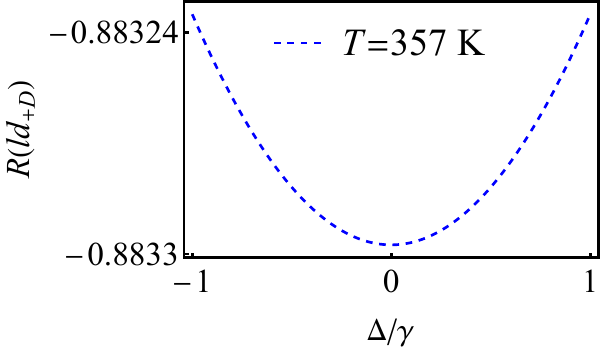} }%
    \hfill
\subfloat[\label{d_2_imag}\centering ]{\includegraphics[width=0.45\columnwidth]{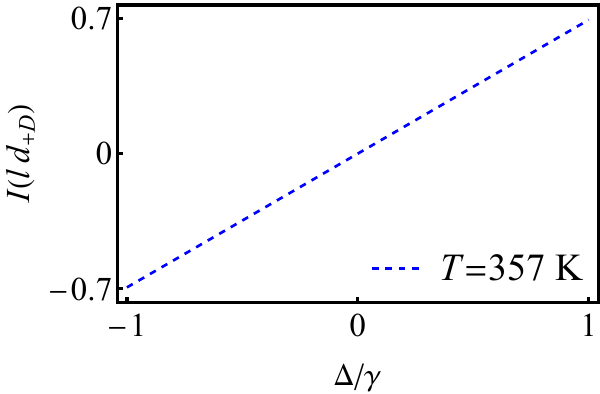} }%
\caption{Variation of (\ref{d_2_real}) real part of $ld_+$ and (\ref{d_2_imag}) imaginary part of $ld_+$  as a function of $\Delta$ for $T=357$ K. Simulation parameters: Mass of a Rb atom, $m=1.44\times 10^{-25}$ Kg, $k_B=1.38\times 10^{-23}$ J/K, velocity of light in vacuum $c=3\times 10^{8}$ m/s, all the other simulation parameters are same as Figure. \ref{d_1l0_real}.}%
    \label{d_1l0-dop}%
\end{figure}

\begin{figure}[hbt!]
    \centering
{\includegraphics[width=0.6\columnwidth]{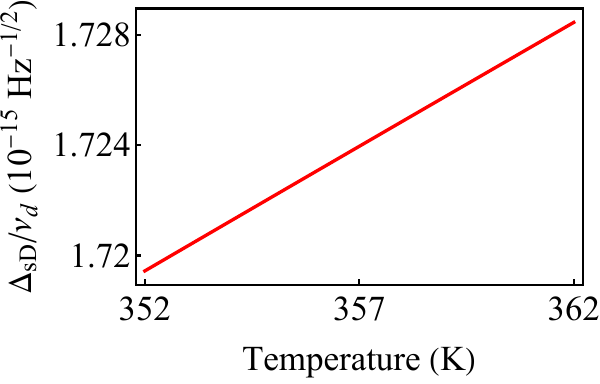}}
\caption{Variation of OAC stability as a function of temperature. $\nu_d/2\pi=3.84\times 10^{14}$ Hz, all the simulation parameters are same as \ref{d_1l0-dop}.}%
    \label{fig5}%
\end{figure}

In this section, we study the effect of temperature on the clock's stability. The atoms in the vapor cell move randomly in all directions with different velocities because of the kinetic energy provided by the thermal environment. The Doppler effect shifts the resonance frequency $\omega_{xy}$, where $\omega_{xy}=\omega_x-\omega_y$, of the moving atom with respect to the incoming laser fields as $\omega_{xy}(1-u/c)$. Where $u$ is the instantaneous velocity of the atom.
This leads to thermal broadening which can be accounted by using Maxwell-Bolzmann distribution \cite{Finkelstein_2023}, as
\begin{equation}\label{dop1}
\begin{split}
&d_{\pm D}=\sqrt{\frac{\alpha}{\pi}}\int\limits_{-\infty}^{\infty}d\Delta_{xyu}\;d_{\pm}(\Delta,\Delta_{xyu}) e^{-\alpha\Delta_{xyu}^2},
\end{split}
\end{equation}
where $d_{\pm D}$ is the thermal broadened $d_{\pm}$, $\alpha=mc^2/2k_BT\omega_{xy}^2$, with $m$ is the mass of a Rb atom, $k_B$ is the Boltzmann constant, $T$ is the temperature, and $\Delta_{xyu}=-\omega_{xy}u/c$. It is not possible to
solve Eq. \eqref{dop1} analytically, therefore, Eq. \eqref{dop1} is numerically integrated, and the corresponding results are shown in Figure. \ref{d_2_real}, Figure. \ref{d_2_imag}, and Figure. \ref{fig5}.
Plots in Figure. \ref{d_2_real} and Figure. \ref{d_2_imag} show the $\mathcal{R}(ld_{+D})$ and $\mathcal{I}(ld_{+D})$ as a function of $\Delta$, respectively, for $\Delta=\Delta_d\lessapprox\gamma$ at $T=357$ K. The blue dashed lines are plotted at temperature 357 K and are slightly broadened than the plots without Doppler broadening (see Figure. \ref{d_1l0}). This implies that the atomic medium is robust against to thermal changes. This is because \cite{Finkelstein_2023} of the collinearly propagating laser fields are always on two-photon resonance, despite of the random atomic thermal motion, with the atoms in a double lambda configuration. The stability $\Delta_{sD}/\nu_d$ of the clock after considering the Doppler broadening is given by Eq. \eqref{delta-sen-3} and Eq. \eqref{dop1} as
\begin{equation}\label{stability-Doppler}
    \frac{\Delta_{sD}}{\nu_d}=\frac{\sqrt{1+4e^{-2ld_{+rD}}}}{\nu_d l\left(d_{+1D}+d_{+2D}\right)\sqrt{|\bar{E}|^2}},
\end{equation}
where $\Delta_{sD},\;d_{+rD},\;d_{+1D}$, and $d_{+2D}$ are the Doppler broadened versions of $\Delta_s,\;d_{+r},\;d_{+1}$, and $d_{+2}$, respectively. We obtained $d_{+r},\;d_{+1D},$ and $d_{+2D}$ by applying Doppler broadening on $d_+$ (Eq. \eqref{stability-Doppler})
The stability of the clock as a function of temperature is plotted in Figure. \ref{fig5} using Eq. \eqref{stability-Doppler}. Note that we assumed the number density of Rb is same (as the number density fluctuate very small amount between 352 K to 362 K) in Figure. \ref{fig5} for simplicity. The stability of the clock changes from $1.719\times 10^{-15}\sqrt{\mbox{Hz}^{-1}}$ to $1.728\times 10^{-15}\sqrt{\mbox{Hz}^{-1}}$ over a temperature range of 352 K to 362 K. Hence the clock is robust against small thermal fluctuations.
\subsection{Collisional broadening}\label{collision}
Collisional broadening is another consequence of the temperature-induced random motion of the atoms. The randomly moving atoms collide within themselves and with the vapor cell walls, leading to decoherence in the form of incoherent pumping between degenerate atomic levels $|b\rangle$ and $|c\rangle$. The effect of collisional pumping is modeled by
adding $\gamma_{cl}(\bar{\sigma}_{cc}-\bar{\sigma}_{bb})$ and $\gamma_{cl}(\bar{\sigma}_{bb}-\bar{\sigma}_{cc})$, with $\gamma_{cl}$ being the collisional pumping rate, terms to the RHS of $\dot{\bar{\sigma}}_{bb}$ and $\dot{\bar{\sigma}}_{cc}$ equations of motion, respectively. 
The effect of collision in presence of thermal Doppler shift is accounted by writing
 $\Gamma=i[\Delta+\Delta_{abu}]-\gamma-\gamma_{cl}/2$, $\Gamma_{d}=i[\Delta+\Delta_{dbu}]-\gamma_d-\gamma_{cl}/2$, $\Gamma_{bc}=-\gamma_{bc}-\gamma_{cl}$.
 Similarly, the effect of collisions in VC2 is accounted by re-writing $\Gamma'=i[\Delta+\Delta_{abu}]-\gamma_d-\gamma_{cl}/2$, $\Gamma'_{d}=i[\Delta+\Delta_{dbu}]-\gamma-\gamma_{cl}/2$, $\Gamma'_{bc}=\Gamma_{bc}$. 
Figure. \ref{dop} shows the clock's stability as a function of temperature with and
without collisions. We assumed that $\gamma_{cl}=\gamma_{bc}$ \cite{PhysRevA.103.022826} in Figure. \ref{dop}. Collisions leads to slight reduction in the clock stability.
\subsection{Laser linewidth}\label{linewidth}
Every laser has a finite linewidth $\zeta$, which adds noise to the atom-laser interactions. The effect of linewidth \cite{McKinstrie:21} is given by random phase $\phi(t)$ variation \cite{Scully_Zubairy_1997}. As all the lasers in the clock design are derived from a single laser, the random in all the laser fields is given by the temporal variation of $\phi$ \cite{PhysRevA.70.053802}. This leads to an additional term $i\dot{\phi}\bar{\sigma}_+$ and $i\dot{\phi}\bar{\sigma}_-$  to the right hand side of Eq. \eqref{sigmapl} and Eq. \eqref{sigmami}, respectively. Due to laser linewidth induced decoherence we re-write $\Gamma=-i[\Delta+\Delta_{abu}]-\gamma-\gamma_{cl}/2-\zeta$ and $\Gamma_d=-i[\Delta+\Delta_{dbu}]-\gamma_d-\gamma_{cl}/2-\zeta$ in MZI-1.

\begin{figure}[hbt!]
    \centering
\subfloat[\label{dop}\centering ]{\includegraphics[width=0.45\columnwidth]{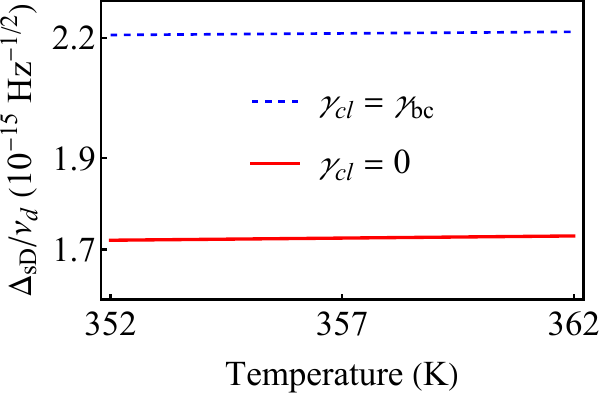} }%
    \qquad
\subfloat[\label{figlin}\centering ]{\includegraphics[width=0.45\columnwidth]{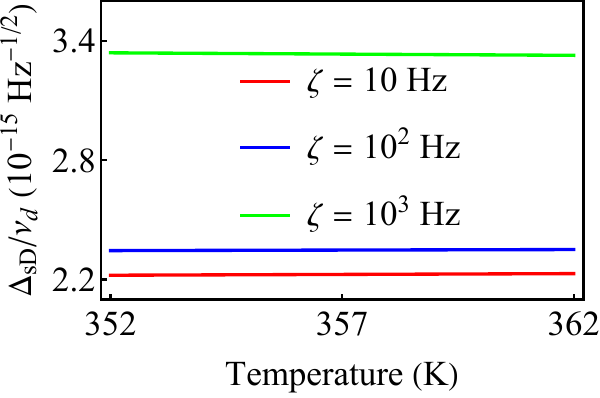} }%
    \caption{(a) Variation of OAC stability as a function of temperature for different $\gamma_{cl}$. (b). Variation of OAC stability as a function of temperature for different linewidth $\zeta$ at $\gamma_{cl}=\gamma_{bc}$ and $\nu_d/2\pi=3.84\times 10^{14}$ Hz. All the other simulation parameters are same as Figure. \ref{d_1l0-dop}.}%
    \label{col-lw}%
\end{figure}
Using the properties $\langle\dot{\phi}\rangle=0$ and $\langle\dot{\phi}(t)\dot{\phi}(t_1)\rangle=2\zeta\delta(t-t_1)$, the phase noise from the $\hat{a}_{+l}$ and $\hat{a}_{-l}$ in the atom-laser interactions is given as
\begin{equation}\label{lw-eq1}
\begin{split}
&\frac{2\zeta(\gamma+\gamma_d+\gamma_{da})^2|\bar{a}_{+0}|^2 e^{2ld_{+r}}(1-e^{ld_{+r}})^2}{[(-\gamma-\gamma_{cl}/2-\zeta)(\gamma+\gamma_d+\gamma_{da})+2|\Omega|^2]^2},\quad\mbox{and}\quad\frac{2\zeta|\bar{a}_{-0}|^2e^{2ld_{-r}}(1-e^{ld_{-r}})^2}{(\gamma+\gamma_{cl}/2+\zeta)^2},
\end{split}
\end{equation}
respectively. Where $d_{-r}$ is real part of $d_-$. We made similar appropriate modifications corresponding to the atom-laser interaction in MZI-2. The plot in Figure. \ref{figlin} shows the stability of the clock as a function of temperature for different values of $\zeta$. The stability of the OAC quickly reduces from $2.2\times 10^{-15}\,\sqrt{\text{Hz}^{-1}}$ to $3.3\times 10^{-15}\,\sqrt{\text{Hz}^{-1}}$ as the $\zeta$ changes from $10\,$Hz to $10^3\,$Hz \cite{10330073} at 357 K temperature. This further enhances the portability and feasibility of our design.
\section{Discussion}\label{Discussion}
In this paragraph, we discuss the advantages of this work over the existing OACs. All the existing OACs use a OFC to downconvert the optical signal to radio frequency for timekeeping. The generated OFC also needs to be calibrated for precise timekeeping. The OAC in this work eliminated the OFC and, consequently, its calibration by using dual interferometers. The Figure. \ref{fig5} indicates that the clock stability remains almost the same for a temperature of $357\pm5\ $K. This proves that the OAC stability is not affected significantly for small thermal variations between VC1 and VC2. Note that the MZI-2 doesn't need to be identical to MZI-1. It is enough if the classical properties (like the length of arms, the length of vapor cells) of MZI-2 and MZI-1 are similar. The Figure. \ref{fig3} shows that small discrepancies in vapor cell length are negligible around the optimum condition $4|\Omega|^2=\kappa\gamma_{bc}l$. In short, the complexity added by the second interferometer in Figure. \ref{fig2} is significantly less than the complexity of OFC and its calibration setup in the existing OAC design. The OFC's femtosecond laser system consumes more energy than the single laser used in Figure. \ref{fig2}. Hence, the OAC described in this work is more portable and energy efficient. The energy efficiency is critical for portable clocks when deployed in outer space.

\par This paragraph compares and contrasts our technique with a few established OAC techniques. Modulation transfer spectroscopy (MTS)~\cite{Zhang:03} is a well-known method often discussed for building frequency standards~\cite{PhysRevLett.87.270801,NEVSKY2001263,Roslund2024}. The MTS ensures that the laser is always locked to atomic resonance~\cite{Lurie:11,Perrella:13}. However, the slope of the MTS signal changes if there is an error in the modulator (AOM, EOM, or EOFS) functioning. As the slope of the signal is directly proportional to the sensitivity and stability of the OAC, an improperly functioning modulator deteriorates the OAC stability. The technique described in this article corrects the laser frequency as well as FS malfunction (see section~\ref{eofs}). This ability of the OAC to self-correct the modulator malfunction is vital when the OAC is deployed in space-based applications. 
\par This paragraph will discuss the OAC based on the lambda system \cite{Yudin:17,Fang2021}. Generally, the lambda system clock stability is proportional to the lower Zeeman level separation, which is the order of radio frequency. Hence, the stability of the clock is often expected to be on par with that of microwave clocks. The key point here is that the lambda system based clocks signal is proportional to the two photon detuning. Since the OAC in this paper is based on dual interferometer technique using a double-lambda configuration, its stability is given by $\nu_d$ (see Eq. \eqref{delta-sen-3}). However, the signal from a single interferometer is linearly dependent on $\Delta$ and $\Delta_d$. In Section \ref{eofs} describes how we can set $\Delta=\Delta_d$ by adjusting the modulating frequency of the FS until we obtain $\mathcal{S}=\mathcal{S}'$. Once $\Delta=\Delta_d$ is enforced, the interferometer output scales as $\Delta/\gamma_d$ (see the imaginary part in the exponential of Eq. \eqref{new-1}, and Eq. \eqref{new-2} for $\gamma=\gamma_d$). Hence, the signal in the double-lambda system-based clock discussed in this paper is proportional to single photon detuning $\Delta$. Whereas, the signal in the lambda system-based clocks is proportional to two-photon detuning. As a result, stability in Eq. \eqref{delta-sen-3} is determined by laser frequency $\nu_d$ and not by radio frequency $\nu-\nu_d$.

\par Once $\Delta=\Delta_d$ is enforced, the signal at the interferometer becomes proportional to $\Delta$. Then, the phase of the interferometer signal is used to implement $\Delta=\Delta_d\approx 0$. This means that we have a simultaneously calibrated $\nu_d$  and $\nu$ to $\ket d'-\ket b'$ and $\ket a-\ket b$ transitions. 
However, since the stability of the OAC is determined by the smallest of the calibrated frequencies, the stability in Eq. \eqref{delta-sen-3} is governed by $\nu_d$ (noting that $\nu_d < \nu$).
\par The two strong fields trap the population in lower levels $\ket b$ and $\ket c$; there will be a tiny amount of population in $\ket d$ also because of non-zero $\gamma_{bc}$. Population from $\ket d$ can decay to an external level outside the double lambda scheme considered in the Figure. \ref{fig1}. However, this can be rectified by pumping the population back to $\ket d$ as usually done by experimenters. This adds some decoherence to the OAC, but it doesn't affect the stability significantly. Theoretically, the decay to the external level can be modeled using a bi-directional incoherent pump from $\ket d$ to the external level $\ket x$. This can be accounted by adding $-p\hat\sigma_{dd}+p\hat\sigma_{xx}$ to Eq. \eqref{eq-sigma-dd} in the appendix. \ref{appendix}, with $p$ as pumping rate, and population conservation equation becomes $\hat\sigma_{aa}+\hat\sigma_{bb}+\hat\sigma_{cc}+\hat\sigma_{dd}+\hat\sigma_{xx}=\hat{1}$, where $\hat{1}$ is the identity matrix. The effect of $p$ on different decoherence terms is accounted as $\Gamma=-i[\Delta+\Delta_{abu}]-\gamma-\gamma_{cl}/2-\zeta-p/2$ and $\Gamma_d=-i[\Delta+\Delta_{dbu}]-\gamma_d-\gamma_{cl}/2-\zeta-p/2$ and $\Gamma_{da}=i(\Delta_{dbu}-\Delta_{abu})-\gamma-\gamma_d-p/2$. A similar appropriate modification has also been made for the MZI-2.
\begin{figure}[hbt!]
    \centering
\includegraphics[width=0.5\linewidth]{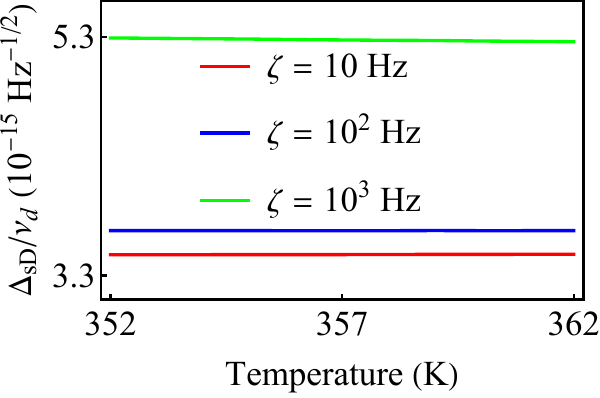}
    \caption{Variation of clock stability as a function of temperature for different $\zeta$ at $p=\gamma$. The stability at 357 K temperature  is reduced to $5.3\times 10^{-15}\sqrt{\mbox{Hz}^{-1}}$ for 1 KHz linewidth laser and $\gamma_{cl}=\gamma_{bc}$. All the other simulation parameters are the same as in the Figure. \ref{col-lw}.}
    \label{fig13}
\end{figure}
\section{Conclusion}
Using four-wave mixing phenomena in alkali metal vapor in a cell, we design a thermally robust portable OAC with improved stability and portability. The portability of the OAC is improved by eliminating the OFC while the one-second stability is improved by an order of magnitude using a new dual interferometer technique. The OFC is eliminated by simultaneously calibrating two laser frequencies and using them to create a beat note at radio frequency for timekeeping purposes. The sensitivity and stability of the OAC are improved by employing a new dual interferometer technique. The double lambda configuration with collinear propagating laser fields renders the OAC robust against thermal noise for room temperature operation. Thus, this OAC does not require complex optical cooling setups. All these factors together lead to the design of a portable
OAC with a standard quantum limited fractional frequency stability of $1.3\times10^{-15}\sqrt{\mbox{Hz}^{-1}}$. After considering broadening effects due to $357\,$K temperature and collisions, the optimum stability of the OAC is reduced to $3.3\times10^{-15}\sqrt{\mbox{Hz}^{-1}}$ for a laser with $1\,$KHz linewidth. As lasers with linewidth in the order of $1\,$KHz are commercially available, this further enhances the portability and feasibility of the technique described in this work. We calculated the sensitivity and stability using the $\ce{^{87}_{}Rb}$ D2 line, but the proposed method can be used for any four-level system showing the interaction as shown in Figure. \ref{fig2}.

\section*{Appendix}\label{appendix}
The equations of motion for the atom-laser interaction shown in Figure. 1 are
\begin{equation}\label{Eqap-1}
\begin{split}
\dot{\hat{\sigma}}_{ba}(z)=\Gamma_{ba}\hat{\sigma}_{ba}(z)+ig_c\hat{a}_1\left(\hat{\sigma}_{aa}-\hat{\sigma}_{bb}\right)
-ig_c\hat{a}_2\hat{\sigma}_{bc}
+i\Omega\hat{\sigma}_{da}(z)+\hat{F}_{ba}(z),
\end{split}
\end{equation}
\begin{equation}
\begin{split}
\dot{\hat{\sigma}}_{ca}(z)=\Gamma_{ca}\hat{\sigma}_{ca}(z)+ig_c\left(\hat{\sigma}_{aa}-\hat{\sigma}_{cc}\right)\hat{a}_2-ig_c\hat{\sigma}_{cb}\hat{a}_1+i\Omega\hat{\sigma}_{da}(z)+\hat{F}_{ca}(z),
\end{split}
\end{equation}
\begin{equation}
\begin{split}
\dot{\hat{\sigma}}_{bc}=\Gamma_{bc}\hat{\sigma}_{bc}-ig_c\hat{a}_2^\dagger\hat{\sigma}_{ba}(z)+ig_c\hat{\sigma}_{ac}(z)\hat{a}_1+i
\Omega \hat{\sigma}_{dc}-i\Omega^* \hat{\sigma}_{bd}+\hat{F}_{bc},
\end{split}
\end{equation}
\begin{equation}
\begin{split}
\dot{\hat{\sigma}}_{db}=\Gamma_{db}\hat{\sigma}_{db}+i\Omega^*\left(\hat{\sigma}_{bb}-\hat{\sigma}_{dd}\right)+i\Omega^*\hat{\sigma}_{cb}-ig_c\hat{a}_1^\dagger\hat{\sigma}_{da}(z)
+\hat{F}_{db},
\end{split}
\end{equation}
\begin{equation}
\begin{split}
\dot{\hat{\sigma}}_{dc}=\Gamma_{dc}\hat{\sigma}_{dc}+i\Omega^*\left(\hat{\sigma}_{cc}-\hat{\sigma}_{dd}\right)+i\Omega^*\hat{\sigma}_{bc}-ig_c\hat{a}_2^\dagger\hat{\sigma}_{da}(z)+\hat{F}_{dc},
\end{split}
\end{equation}
\begin{equation}
\begin{split}
\dot{\hat{\sigma}}_{da}(z)=\Gamma_{da}\hat{\sigma}_{da}(z)-ig_c\hat{\sigma}_{db}\hat{a}_1-ig_c\hat{\sigma}_{dc}\hat{a}_2+i\Omega^*\hat{\sigma}_{ba}(z)+i\Omega^*\hat{\sigma}_{ca}(z)+\hat{F}_{da}(z),
\end{split}
\end{equation}
\begin{equation}
\begin{split}
\dot{\hat{\sigma}}_{aa}=ig_c\hat{a}_1^\dagger\hat{\sigma}_{ba}(z)-ig_c\hat{\sigma}_{ab}(z)\hat{a}_1+ig_c\hat{a}_2^\dagger\hat{\sigma}_{ca}(z)-ig_c\hat{\sigma}_{ac}(z)\hat{a}_2-2\gamma\hat{\sigma}_{aa}+\hat{F}_{aa},
\end{split}
\end{equation}
\begin{equation}\label{eq-sigma-dd}
\begin{split}
\dot{\hat{\sigma}}_{dd}=i\Omega^{*}\hat{\sigma}_{bd}-i\hat{\sigma}_{db}\Omega+i\Omega^{*}\hat{\sigma}_{cd}-i\hat{\sigma}_{dc}\Omega-2\gamma_d\hat{\sigma}_{dd}+\hat{F}_{dd},
\end{split}
\end{equation}
\begin{equation}
\begin{split}
\dot{\hat{\sigma}}_{bb}=i\hat{\sigma}_{db}\Omega-i\Omega^{*}\hat{\sigma}_{bd}+ig_c\hat{\sigma}_{ab}(z)\hat{a}_1
-ig_c\hat{a}_1^\dagger\hat{\sigma}_{ba}(z)+\gamma_d\hat{\sigma}_{dd}+\gamma\hat{\sigma}_{aa}+\hat{F}_{bb},
\end{split}
\end{equation}
\begin{equation}\label{Eqap-10}
\begin{split}
\dot{\hat{\sigma}}_{cc}=i\hat{\sigma}_{dc}\Omega-i\Omega^{*}\hat{\sigma}_{cd}+ig_c\hat{\sigma}_{ac}(z)\hat{a}_2-ig_c\hat{a}_2^\dagger\hat{\sigma}_{ca}(z)+\gamma_d\hat{\sigma}_{dd}+\gamma\hat{\sigma}_{aa}+\hat{F}_{cc},
\end{split}
\end{equation}
As the quantum fields $\hat{a}_1$ and $\hat{a}_2$ are weak, their dynamics are given by first order perturbation. The equations of motion are given as
\begin{equation}
\begin{split}
&\left(\frac{\partial}{\partial t}+c\frac{\partial}{\partial z}\right)\hat{a}_1=-\frac{iNcg_c}{l}\hat{\sigma}_{ba}^{(1)};\quad
\left(\frac{\partial}{\partial t}+c\frac{\partial}{\partial z}\right)\hat{a}_2=-\frac{iNcg_c}{l}\hat{\sigma}_{ca}^{(1)},
\end{split}
\end{equation}
\begin{equation}
\begin{split}
\dot{\hat{\sigma}}_{ba}^{(1)}(z)=\Gamma_{ba}\hat{\sigma}_{ba}^{(1)}(z)+ig_c\hat{a}_1\left(\bar{\sigma}_{aa}-\bar{\sigma}_{bb}\right)-ig_c\hat{a}_2\bar{\sigma}_{bc}
+i\Omega\hat{\sigma}_{da}^{(1)}(z)+\hat{F}_{ba}(z),
\end{split}
\end{equation}
\begin{equation}
\begin{split}
\dot{\hat{\sigma}}_{ca}^{(1)}(z)=\Gamma_{ca}\hat{\sigma}_{ca}{(1)}(z)+ig_c\left(\bar{\sigma}_{aa}-\bar{\sigma}_{cc}\right)\hat{a}_2-ig_c\bar{\sigma}_{cb}\hat{a}_1
+i\Omega\hat{\sigma}_{da}^{(1)}(z)+\hat{F}_{ca}(z),
\end{split}
\end{equation}
\begin{equation}
\begin{split}
\dot{\hat{\sigma}}_{da}^{(1)}(z)=\Gamma_{da}\hat{\sigma}_{da}^{(1)}(z)-ig_c\bar{\sigma}_{db}\hat{a}_1-ig_c\bar{\sigma}_{dc}\hat{a}_2
+i\Omega^*\hat{\sigma}_{ba}^{(1)}(z)+i\Omega^*\hat{\sigma}_{ca}^{(1)}(z)+\hat{F}_{da}(z).
\end{split}
\end{equation}
The zero order equations of motion are given as
\begin{equation}\label{zero-1}
\begin{split}
\dot{\bar{\sigma}}_{bc}=\Gamma_{bc}\bar{\sigma}_{bc}+i
\Omega \bar{\sigma}_{dc}-i\Omega^* \bar{\sigma}_{bd},
\end{split}
\end{equation}
\begin{equation}
\begin{split}
\dot{\bar{\sigma}}_{db}=\Gamma_{db}\bar{\sigma}_{db}+i\Omega^*\left(\bar{\sigma}_{bb}-\bar{\sigma}_{dd}\right)+i\Omega^*\bar{\sigma}_{cb},
\end{split}
\end{equation}
\begin{equation}
\begin{split}
\dot{\bar{\sigma}}_{dc}=\Gamma_{dc}\bar{\sigma}_{dc}+i\Omega^*\left(\bar{\sigma}_{cc}-\bar{\sigma}_{dd}\right)+i\Omega^*\bar{\sigma}_{bc},
\end{split}
\end{equation}
\begin{equation}
\begin{split}
\dot{\bar{\sigma}}_{aa}=-2\gamma\bar{\sigma}_{aa},
\end{split}
\end{equation}
\begin{equation}
\begin{split}
\dot{\bar{\sigma}}_{dd}=i\Omega^{*}\bar{\sigma}_{bd}-i\bar{\sigma}_{db}\Omega+i\Omega^{*}\bar{\sigma}_{cd}-i\bar{\sigma}_{dc}\Omega-2\gamma_d\bar{\sigma}_{dd},
\end{split}
\end{equation}
\begin{equation}
\begin{split}
\dot{\bar{\sigma}}_{bb}=i\bar{\sigma}_{db}\Omega-i\Omega^{*}\bar{\sigma}_{bd}+\gamma_d\bar{\sigma}_{dd}+\gamma\bar{\sigma}_{aa},
\end{split}
\end{equation}
\begin{equation}\label{zero-7}
\begin{split}
\dot{\bar{\sigma}}_{cc}=i\bar{\sigma}_{dc}\Omega-i\Omega^{*}\bar{\sigma}_{cd}+\gamma_d\bar{\sigma}_{dd}+\gamma\bar{\sigma}_{aa}.
\end{split}
\end{equation}
The non-zero correlations between the atomic noise operators are
\begin{equation*}
\begin{split}
   \langle \hat{F}_{ba}(z,t)\hat{F}^\dagger_{ba}(z_1,t_1)\rangle=\frac{l}{N}[-(\Gamma_{ab}+\Gamma_{ba})\bar{\sigma}_{bb}+\gamma\bar{\sigma}_{aa}+\gamma_d\bar{\sigma}_{dd}]
   \delta(z-z_1)\delta(t-t_1),
   \end{split}
\end{equation*}
\begin{equation*}
\begin{split}
 \langle\hat{F}_{ba}(z,t)\hat{F}^\dagger_{da}(z_1,t_1)\rangle=\frac{l}{N}[-(\Gamma_{ad}+\Gamma_{ba}-\Gamma_{bd})\bar{\sigma}_{bd}]
 \delta(z-z_1)\delta(t-t_1),  
 \end{split}
\end{equation*}
\begin{equation*}
\begin{split}
  \langle\hat{F}_{ba}(z,t)\hat{F}^\dagger_{ca}(z_1,t_1)\rangle=\frac{l}{N}[-(\Gamma_{ac}+\Gamma_{ba}-\Gamma_{bc})\bar{\sigma}_{bc}]
  \delta(z-z_1)\delta(t-t_1), 
  \end{split}
\end{equation*}
  \begin{equation*}
  \begin{split}
      \langle\hat{F}_{da}(z,t)\hat{F}^\dagger_{da}(z_1,t_1)\rangle=\frac{l}{N}[-(\Gamma_{ad}+\Gamma_{da}+2\gamma_d)\bar{\sigma}_{dd}]
      \delta(z-z_1)\delta(t-t_1),
      \end{split}
  \end{equation*}
  \begin{equation*}
  \begin{split}
     \langle\hat{F}_{da}(z,t)\hat{F}^\dagger_{ca}(z_1,t_1)\rangle=\frac{l}{N}[-(\Gamma_{ac}+\Gamma_{da}-\Gamma_{dc})\bar{\sigma}_{dc}]
     \delta(z-z_1)\delta(t-t_1), 
     \end{split}
  \end{equation*}
  \begin{equation*}
  \begin{split}
    \langle\hat{F}_{ca}(z,t)\hat{F}^\dagger_{ca}(z_1,t_1)\rangle=\frac{l}{N}[-(\Gamma_{ac}+\Gamma_{ca})\bar{\sigma}_{cc}+\gamma\bar{\sigma}_{aa}+\gamma_d\bar{\sigma}_{dd}]
    \delta(z-z_1)\delta(t-t_1).
    \end{split}
  \end{equation*}
 Here, we write the equations of motion given in the atom-laser interaction in Figure. 1. Similarly, we get the equations of motion of the atom-laser interaction in Figure. 2.
\bibliography{apssamp}

\providecommand{\noopsort}[1]{}\providecommand{\singleletter}[1]{#1}%
\begin{thebibliography}{64}%
\makeatletter
\providecommand \@ifxundefined [1]{%
 \@ifx{#1\undefined}
}%
\providecommand \@ifnum [1]{%
 \ifnum #1\expandafter \@firstoftwo
 \else \expandafter \@secondoftwo
 \fi
}%
\providecommand \@ifx [1]{%
 \ifx #1\expandafter \@firstoftwo
 \else \expandafter \@secondoftwo
 \fi
}%
\providecommand \natexlab [1]{#1}%
\providecommand \enquote  [1]{``#1''}%
\providecommand \bibnamefont  [1]{#1}%
\providecommand \bibfnamefont [1]{#1}%
\providecommand \citenamefont [1]{#1}%
\providecommand \href@noop [0]{\@secondoftwo}%
\providecommand \href [0]{\begingroup \@sanitize@url \@href}%
\providecommand \@href[1]{\@@startlink{#1}\@@href}%
\providecommand \@@href[1]{\endgroup#1\@@endlink}%
\providecommand \@sanitize@url [0]{\catcode `\\12\catcode `\$12\catcode
  `\&12\catcode `\#12\catcode `\^12\catcode `\_12\catcode `\%12\relax}%
\providecommand \@@startlink[1]{}%
\providecommand \@@endlink[0]{}%
\providecommand \url  [0]{\begingroup\@sanitize@url \@url }%
\providecommand \@url [1]{\endgroup\@href {#1}{\urlprefix }}%
\providecommand \urlprefix  [0]{URL }%
\providecommand \Eprint [0]{\href }%
\providecommand \doibase [0]{https://doi.org/}%
\providecommand \selectlanguage [0]{\@gobble}%
\providecommand \bibinfo  [0]{\@secondoftwo}%
\providecommand \bibfield  [0]{\@secondoftwo}%
\providecommand \translation [1]{[#1]}%
\providecommand \BibitemOpen [0]{}%
\providecommand \bibitemStop [0]{}%
\providecommand \bibitemNoStop [0]{.\EOS\space}%
\providecommand \EOS [0]{\spacefactor3000\relax}%
\providecommand \BibitemShut  [1]{\csname bibitem#1\endcsname}%
\let\auto@bib@innerbib\@empty
\bibitem [{\citenamefont {Ludlow}\ \emph {et~al.}(2015)\citenamefont {Ludlow},
  \citenamefont {Boyd}, \citenamefont {Ye}, \citenamefont {Peik},\ and\
  \citenamefont {Schmidt}}]{RevModPhys.87.637}%
  \BibitemOpen
  \bibfield  {author} {\bibinfo {author} {\bibfnamefont {A.~D.}\ \bibnamefont
  {Ludlow}}, \bibinfo {author} {\bibfnamefont {M.~M.}\ \bibnamefont {Boyd}},
  \bibinfo {author} {\bibfnamefont {J.}~\bibnamefont {Ye}}, \bibinfo {author}
  {\bibfnamefont {E.}~\bibnamefont {Peik}},\ and\ \bibinfo {author}
  {\bibfnamefont {P.~O.}\ \bibnamefont {Schmidt}},\ }\bibfield  {title}
  {\bibinfo {title} {Optical atomic clocks},\ }\href
  {https://doi.org/10.1103/RevModPhys.87.637} {\bibfield  {journal} {\bibinfo
  {journal} {Rev. Mod. Phys.}\ }\textbf {\bibinfo {volume} {87}},\ \bibinfo
  {pages} {637} (\bibinfo {year} {2015})}\BibitemShut {NoStop}%
\bibitem [{\citenamefont {Poli}\ \emph {et~al.}(2013)\citenamefont {Poli},
  \citenamefont {Oates}, \citenamefont {Gill},\ and\ \citenamefont
  {Tino}}]{Poli2013}%
  \BibitemOpen
  \bibfield  {author} {\bibinfo {author} {\bibfnamefont {N.}~\bibnamefont
  {Poli}}, \bibinfo {author} {\bibfnamefont {C.~W.}\ \bibnamefont {Oates}},
  \bibinfo {author} {\bibfnamefont {P.}~\bibnamefont {Gill}},\ and\ \bibinfo
  {author} {\bibfnamefont {G.~M.}\ \bibnamefont {Tino}},\ }\bibfield  {title}
  {\bibinfo {title} {Optical atomic clocks},\ }\href
  {https://doi.org/10.1393/ncr/i2013-10095-x} {\bibfield  {journal} {\bibinfo
  {journal} {La Rivista del Nuovo Cimento}\ }\textbf {\bibinfo {volume} {36}},\
  \bibinfo {pages} {555} (\bibinfo {year} {2013})}\BibitemShut {NoStop}%
\bibitem [{\citenamefont {Roslund}\ \emph {et~al.}(2024)\citenamefont
  {Roslund}, \citenamefont {Cing{\"o}z}, \citenamefont {Lunden}, \citenamefont
  {Partridge}, \citenamefont {Kowligy}, \citenamefont {Roller}, \citenamefont
  {Sheredy}, \citenamefont {Skulason}, \citenamefont {Song}, \citenamefont
  {Abo-Shaeer},\ and\ \citenamefont {Boyd}}]{Roslund2024}%
  \BibitemOpen
  \bibfield  {author} {\bibinfo {author} {\bibfnamefont {J.~D.}\ \bibnamefont
  {Roslund}}, \bibinfo {author} {\bibfnamefont {A.}~\bibnamefont {Cing{\"o}z}},
  \bibinfo {author} {\bibfnamefont {W.~D.}\ \bibnamefont {Lunden}}, \bibinfo
  {author} {\bibfnamefont {G.~B.}\ \bibnamefont {Partridge}}, \bibinfo {author}
  {\bibfnamefont {A.~S.}\ \bibnamefont {Kowligy}}, \bibinfo {author}
  {\bibfnamefont {F.}~\bibnamefont {Roller}}, \bibinfo {author} {\bibfnamefont
  {D.~B.}\ \bibnamefont {Sheredy}}, \bibinfo {author} {\bibfnamefont {G.~E.}\
  \bibnamefont {Skulason}}, \bibinfo {author} {\bibfnamefont {J.~P.}\
  \bibnamefont {Song}}, \bibinfo {author} {\bibfnamefont {J.~R.}\ \bibnamefont
  {Abo-Shaeer}},\ and\ \bibinfo {author} {\bibfnamefont {M.~M.}\ \bibnamefont
  {Boyd}},\ }\bibfield  {title} {\bibinfo {title} {Optical clocks at sea},\
  }\href {https://doi.org/10.1038/s41586-024-07225-2} {\bibfield  {journal}
  {\bibinfo  {journal} {Nature}\ }\textbf {\bibinfo {volume} {628}},\ \bibinfo
  {pages} {736} (\bibinfo {year} {2024})}\BibitemShut {NoStop}%
\bibitem [{\citenamefont {Bandi}(2024)}]{bandi2024comprehensive}%
  \BibitemOpen
  \bibfield  {author} {\bibinfo {author} {\bibfnamefont {T.~N.}\ \bibnamefont
  {Bandi}},\ }\bibfield  {title} {\bibinfo {title} {A comprehensive overview of
  atomic clocks and their applications},\ }\href@noop {} {\bibfield  {journal}
  {\bibinfo  {journal} {Demo Journal}\ }\textbf {\bibinfo {volume} {1}},\
  \bibinfo {pages} {40} (\bibinfo {year} {2024})}\BibitemShut {NoStop}%
\bibitem [{\citenamefont {Schioppo}\ \emph {et~al.}(2017)\citenamefont
  {Schioppo}, \citenamefont {Brown}, \citenamefont {McGrew}, \citenamefont
  {Hinkley}, \citenamefont {Fasano}, \citenamefont {Beloy}, \citenamefont
  {Yoon}, \citenamefont {Milani}, \citenamefont {Nicolodi}, \citenamefont
  {Sherman}, \citenamefont {Phillips}, \citenamefont {Oates},\ and\
  \citenamefont {Ludlow}}]{Schioppo2017}%
  \BibitemOpen
  \bibfield  {author} {\bibinfo {author} {\bibfnamefont {M.}~\bibnamefont
  {Schioppo}}, \bibinfo {author} {\bibfnamefont {R.~C.}\ \bibnamefont {Brown}},
  \bibinfo {author} {\bibfnamefont {W.~F.}\ \bibnamefont {McGrew}}, \bibinfo
  {author} {\bibfnamefont {N.}~\bibnamefont {Hinkley}}, \bibinfo {author}
  {\bibfnamefont {R.~J.}\ \bibnamefont {Fasano}}, \bibinfo {author}
  {\bibfnamefont {K.}~\bibnamefont {Beloy}}, \bibinfo {author} {\bibfnamefont
  {T.~H.}\ \bibnamefont {Yoon}}, \bibinfo {author} {\bibfnamefont
  {G.}~\bibnamefont {Milani}}, \bibinfo {author} {\bibfnamefont
  {D.}~\bibnamefont {Nicolodi}}, \bibinfo {author} {\bibfnamefont {J.~A.}\
  \bibnamefont {Sherman}}, \bibinfo {author} {\bibfnamefont {N.~B.}\
  \bibnamefont {Phillips}}, \bibinfo {author} {\bibfnamefont {C.~W.}\
  \bibnamefont {Oates}},\ and\ \bibinfo {author} {\bibfnamefont {A.~D.}\
  \bibnamefont {Ludlow}},\ }\bibfield  {title} {\bibinfo {title} {Ultrastable
  optical clock with two cold-atom ensembles},\ }\href
  {https://doi.org/10.1038/nphoton.2016.231} {\bibfield  {journal} {\bibinfo
  {journal} {Nature Photonics}\ }\textbf {\bibinfo {volume} {11}},\ \bibinfo
  {pages} {48} (\bibinfo {year} {2017})}\BibitemShut {NoStop}%
\bibitem [{\citenamefont {Bothwell}\ \emph {et~al.}(2019)\citenamefont
  {Bothwell}, \citenamefont {Kedar}, \citenamefont {Oelker}, \citenamefont
  {Robinson}, \citenamefont {Bromley}, \citenamefont {Tew}, \citenamefont
  {Ye},\ and\ \citenamefont {Kennedy}}]{Bothwell_2019}%
  \BibitemOpen
  \bibfield  {author} {\bibinfo {author} {\bibfnamefont {T.}~\bibnamefont
  {Bothwell}}, \bibinfo {author} {\bibfnamefont {D.}~\bibnamefont {Kedar}},
  \bibinfo {author} {\bibfnamefont {E.}~\bibnamefont {Oelker}}, \bibinfo
  {author} {\bibfnamefont {J.~M.}\ \bibnamefont {Robinson}}, \bibinfo {author}
  {\bibfnamefont {S.~L.}\ \bibnamefont {Bromley}}, \bibinfo {author}
  {\bibfnamefont {W.~L.}\ \bibnamefont {Tew}}, \bibinfo {author} {\bibfnamefont
  {J.}~\bibnamefont {Ye}},\ and\ \bibinfo {author} {\bibfnamefont {C.~J.}\
  \bibnamefont {Kennedy}},\ }\bibfield  {title} {\bibinfo {title} {Jila sri
  optical lattice clock with uncertainty of 2.0 $\times 10^{-18}$},\ }\href
  {https://doi.org/10.1088/1681-7575/ab4089} {\bibfield  {journal} {\bibinfo
  {journal} {Metrologia}\ }\textbf {\bibinfo {volume} {56}},\ \bibinfo {pages}
  {065004} (\bibinfo {year} {2019})}\BibitemShut {NoStop}%
\bibitem [{\citenamefont {McGrew}\ \emph {et~al.}(2018)\citenamefont {McGrew},
  \citenamefont {Zhang}, \citenamefont {Fasano}, \citenamefont {Sch{\"a}ffer},
  \citenamefont {Beloy}, \citenamefont {Nicolodi}, \citenamefont {Brown},
  \citenamefont {Hinkley}, \citenamefont {Milani}, \citenamefont {Schioppo},
  \citenamefont {Yoon},\ and\ \citenamefont {Ludlow}}]{McGrew2018}%
  \BibitemOpen
  \bibfield  {author} {\bibinfo {author} {\bibfnamefont {W.~F.}\ \bibnamefont
  {McGrew}}, \bibinfo {author} {\bibfnamefont {X.}~\bibnamefont {Zhang}},
  \bibinfo {author} {\bibfnamefont {R.~J.}\ \bibnamefont {Fasano}}, \bibinfo
  {author} {\bibfnamefont {S.~A.}\ \bibnamefont {Sch{\"a}ffer}}, \bibinfo
  {author} {\bibfnamefont {K.}~\bibnamefont {Beloy}}, \bibinfo {author}
  {\bibfnamefont {D.}~\bibnamefont {Nicolodi}}, \bibinfo {author}
  {\bibfnamefont {R.~C.}\ \bibnamefont {Brown}}, \bibinfo {author}
  {\bibfnamefont {N.}~\bibnamefont {Hinkley}}, \bibinfo {author} {\bibfnamefont
  {G.}~\bibnamefont {Milani}}, \bibinfo {author} {\bibfnamefont
  {M.}~\bibnamefont {Schioppo}}, \bibinfo {author} {\bibfnamefont {T.~H.}\
  \bibnamefont {Yoon}},\ and\ \bibinfo {author} {\bibfnamefont {A.~D.}\
  \bibnamefont {Ludlow}},\ }\bibfield  {title} {\bibinfo {title} {Atomic clock
  performance enabling geodesy below the centimetre level},\ }\href
  {https://doi.org/10.1038/s41586-018-0738-2} {\bibfield  {journal} {\bibinfo
  {journal} {Nature}\ }\textbf {\bibinfo {volume} {564}},\ \bibinfo {pages}
  {87} (\bibinfo {year} {2018})}\BibitemShut {NoStop}%
\bibitem [{\citenamefont {Brewer}\ \emph {et~al.}(2019)\citenamefont {Brewer},
  \citenamefont {Chen}, \citenamefont {Hankin}, \citenamefont {Clements},
  \citenamefont {Chou}, \citenamefont {Wineland}, \citenamefont {Hume},\ and\
  \citenamefont {Leibrandt}}]{PhysRevLett.123.033201}%
  \BibitemOpen
  \bibfield  {author} {\bibinfo {author} {\bibfnamefont {S.~M.}\ \bibnamefont
  {Brewer}}, \bibinfo {author} {\bibfnamefont {J.-S.}\ \bibnamefont {Chen}},
  \bibinfo {author} {\bibfnamefont {A.~M.}\ \bibnamefont {Hankin}}, \bibinfo
  {author} {\bibfnamefont {E.~R.}\ \bibnamefont {Clements}}, \bibinfo {author}
  {\bibfnamefont {C.~W.}\ \bibnamefont {Chou}}, \bibinfo {author}
  {\bibfnamefont {D.~J.}\ \bibnamefont {Wineland}}, \bibinfo {author}
  {\bibfnamefont {D.~B.}\ \bibnamefont {Hume}},\ and\ \bibinfo {author}
  {\bibfnamefont {D.~R.}\ \bibnamefont {Leibrandt}},\ }\bibfield  {title}
  {\bibinfo {title} {$^{27}{\mathrm{al}}^{+}$ quantum-logic clock with a
  systematic uncertainty below ${10}^{\ensuremath{-}18}$},\ }\href
  {https://doi.org/10.1103/PhysRevLett.123.033201} {\bibfield  {journal}
  {\bibinfo  {journal} {Phys. Rev. Lett.}\ }\textbf {\bibinfo {volume} {123}},\
  \bibinfo {pages} {033201} (\bibinfo {year} {2019})}\BibitemShut {NoStop}%
\bibitem [{\citenamefont {Derevianko}\ \emph {et~al.}(2012)\citenamefont
  {Derevianko}, \citenamefont {Dzuba},\ and\ \citenamefont
  {Flambaum}}]{PhysRevLett.109.180801}%
  \BibitemOpen
  \bibfield  {author} {\bibinfo {author} {\bibfnamefont {A.}~\bibnamefont
  {Derevianko}}, \bibinfo {author} {\bibfnamefont {V.~A.}\ \bibnamefont
  {Dzuba}},\ and\ \bibinfo {author} {\bibfnamefont {V.~V.}\ \bibnamefont
  {Flambaum}},\ }\bibfield  {title} {\bibinfo {title} {Highly charged ions as a
  basis of optical atomic clockwork of exceptional accuracy},\ }\href
  {https://doi.org/10.1103/PhysRevLett.109.180801} {\bibfield  {journal}
  {\bibinfo  {journal} {Phys. Rev. Lett.}\ }\textbf {\bibinfo {volume} {109}},\
  \bibinfo {pages} {180801} (\bibinfo {year} {2012})}\BibitemShut {NoStop}%
\bibitem [{\citenamefont {Lange}\ \emph {et~al.}(2021)\citenamefont {Lange},
  \citenamefont {Huntemann}, \citenamefont {Rahm}, \citenamefont {Sanner},
  \citenamefont {Shao}, \citenamefont {Lipphardt}, \citenamefont {Tamm},
  \citenamefont {Weyers},\ and\ \citenamefont {Peik}}]{PhysRevLett.126.011102}%
  \BibitemOpen
  \bibfield  {author} {\bibinfo {author} {\bibfnamefont {R.}~\bibnamefont
  {Lange}}, \bibinfo {author} {\bibfnamefont {N.}~\bibnamefont {Huntemann}},
  \bibinfo {author} {\bibfnamefont {J.~M.}\ \bibnamefont {Rahm}}, \bibinfo
  {author} {\bibfnamefont {C.}~\bibnamefont {Sanner}}, \bibinfo {author}
  {\bibfnamefont {H.}~\bibnamefont {Shao}}, \bibinfo {author} {\bibfnamefont
  {B.}~\bibnamefont {Lipphardt}}, \bibinfo {author} {\bibfnamefont
  {C.}~\bibnamefont {Tamm}}, \bibinfo {author} {\bibfnamefont {S.}~\bibnamefont
  {Weyers}},\ and\ \bibinfo {author} {\bibfnamefont {E.}~\bibnamefont {Peik}},\
  }\bibfield  {title} {\bibinfo {title} {Improved limits for violations of
  local position invariance from atomic clock comparisons},\ }\href
  {https://doi.org/10.1103/PhysRevLett.126.011102} {\bibfield  {journal}
  {\bibinfo  {journal} {Phys. Rev. Lett.}\ }\textbf {\bibinfo {volume} {126}},\
  \bibinfo {pages} {011102} (\bibinfo {year} {2021})}\BibitemShut {NoStop}%
\bibitem [{\citenamefont {Holliman}\ \emph {et~al.}(2022)\citenamefont
  {Holliman}, \citenamefont {Fan}, \citenamefont {Contractor}, \citenamefont
  {Brewer},\ and\ \citenamefont {Jayich}}]{PhysRevLett.128.033202}%
  \BibitemOpen
  \bibfield  {author} {\bibinfo {author} {\bibfnamefont {C.~A.}\ \bibnamefont
  {Holliman}}, \bibinfo {author} {\bibfnamefont {M.}~\bibnamefont {Fan}},
  \bibinfo {author} {\bibfnamefont {A.}~\bibnamefont {Contractor}}, \bibinfo
  {author} {\bibfnamefont {S.~M.}\ \bibnamefont {Brewer}},\ and\ \bibinfo
  {author} {\bibfnamefont {A.~M.}\ \bibnamefont {Jayich}},\ }\bibfield  {title}
  {\bibinfo {title} {Radium ion optical clock},\ }\href
  {https://doi.org/10.1103/PhysRevLett.128.033202} {\bibfield  {journal}
  {\bibinfo  {journal} {Phys. Rev. Lett.}\ }\textbf {\bibinfo {volume} {128}},\
  \bibinfo {pages} {033202} (\bibinfo {year} {2022})}\BibitemShut {NoStop}%
\bibitem [{\citenamefont {Zheng}\ \emph {et~al.}(2022)\citenamefont {Zheng},
  \citenamefont {Dolde}, \citenamefont {Lochab}, \citenamefont {Merriman},
  \citenamefont {Li},\ and\ \citenamefont {Kolkowitz}}]{Zheng2022}%
  \BibitemOpen
  \bibfield  {author} {\bibinfo {author} {\bibfnamefont {X.}~\bibnamefont
  {Zheng}}, \bibinfo {author} {\bibfnamefont {J.}~\bibnamefont {Dolde}},
  \bibinfo {author} {\bibfnamefont {V.}~\bibnamefont {Lochab}}, \bibinfo
  {author} {\bibfnamefont {B.~N.}\ \bibnamefont {Merriman}}, \bibinfo {author}
  {\bibfnamefont {H.}~\bibnamefont {Li}},\ and\ \bibinfo {author}
  {\bibfnamefont {S.}~\bibnamefont {Kolkowitz}},\ }\bibfield  {title} {\bibinfo
  {title} {Differential clock comparisons with a multiplexed optical lattice
  clock},\ }\href {https://doi.org/10.1038/s41586-021-04344-y} {\bibfield
  {journal} {\bibinfo  {journal} {Nature}\ }\textbf {\bibinfo {volume} {602}},\
  \bibinfo {pages} {425} (\bibinfo {year} {2022})}\BibitemShut {NoStop}%
\bibitem [{\citenamefont {Yu}\ \emph {et~al.}(2024)\citenamefont {Yu},
  \citenamefont {Zhang}, \citenamefont {Zhang},\ and\ \citenamefont
  {Chen}}]{lasing}%
  \BibitemOpen
  \bibfield  {author} {\bibinfo {author} {\bibfnamefont {D.}~\bibnamefont
  {Yu}}, \bibinfo {author} {\bibfnamefont {J.}~\bibnamefont {Zhang}}, \bibinfo
  {author} {\bibfnamefont {S.}~\bibnamefont {Zhang}},\ and\ \bibinfo {author}
  {\bibfnamefont {J.}~\bibnamefont {Chen}},\ }\bibfield  {title} {\bibinfo
  {title} {Prospects for an active optical clock based on cavityless lasing},\
  }\href {https://doi.org/https://doi.org/10.1002/qute.202300308} {\bibfield
  {journal} {\bibinfo  {journal} {Advanced Quantum Technologies}\ }\textbf
  {\bibinfo {volume} {7}},\ \bibinfo {pages} {2300308} (\bibinfo {year}
  {2024})}\BibitemShut {NoStop}%
\bibitem [{\citenamefont {Ushijima}\ \emph {et~al.}(2015)\citenamefont
  {Ushijima}, \citenamefont {Takamoto}, \citenamefont {Das}, \citenamefont
  {Ohkubo},\ and\ \citenamefont {Katori}}]{Ushijima2015}%
  \BibitemOpen
  \bibfield  {author} {\bibinfo {author} {\bibfnamefont {I.}~\bibnamefont
  {Ushijima}}, \bibinfo {author} {\bibfnamefont {M.}~\bibnamefont {Takamoto}},
  \bibinfo {author} {\bibfnamefont {M.}~\bibnamefont {Das}}, \bibinfo {author}
  {\bibfnamefont {T.}~\bibnamefont {Ohkubo}},\ and\ \bibinfo {author}
  {\bibfnamefont {H.}~\bibnamefont {Katori}},\ }\bibfield  {title} {\bibinfo
  {title} {Cryogenic optical lattice clocks},\ }\href
  {https://doi.org/10.1038/nphoton.2015.5} {\bibfield  {journal} {\bibinfo
  {journal} {Nature Photonics}\ }\textbf {\bibinfo {volume} {9}},\ \bibinfo
  {pages} {185} (\bibinfo {year} {2015})}\BibitemShut {NoStop}%
\bibitem [{\citenamefont {Chou}\ \emph
  {et~al.}(2010{\natexlab{a}})\citenamefont {Chou}, \citenamefont {Hume},
  \citenamefont {Rosenband},\ and\ \citenamefont
  {Wineland}}]{doi:10.1126/science.1192720}%
  \BibitemOpen
  \bibfield  {author} {\bibinfo {author} {\bibfnamefont {C.~W.}\ \bibnamefont
  {Chou}}, \bibinfo {author} {\bibfnamefont {D.~B.}\ \bibnamefont {Hume}},
  \bibinfo {author} {\bibfnamefont {T.}~\bibnamefont {Rosenband}},\ and\
  \bibinfo {author} {\bibfnamefont {D.~J.}\ \bibnamefont {Wineland}},\
  }\bibfield  {title} {\bibinfo {title} {Optical clocks and relativity},\
  }\href {https://doi.org/10.1126/science.1192720} {\bibfield  {journal}
  {\bibinfo  {journal} {Science}\ }\textbf {\bibinfo {volume} {329}},\ \bibinfo
  {pages} {1630} (\bibinfo {year} {2010}{\natexlab{a}})}\BibitemShut {NoStop}%
\bibitem [{\citenamefont {Newman}\ \emph {et~al.}(2019)\citenamefont {Newman},
  \citenamefont {Maurice}, \citenamefont {Drake}, \citenamefont {Stone},
  \citenamefont {Briles}, \citenamefont {Spencer}, \citenamefont {Fredrick},
  \citenamefont {Li}, \citenamefont {Westly}, \citenamefont {Ilic},
  \citenamefont {Shen}, \citenamefont {Suh}, \citenamefont {Yang},
  \citenamefont {Johnson}, \citenamefont {Johnson}, \citenamefont {Hollberg},
  \citenamefont {Vahala}, \citenamefont {Srinivasan}, \citenamefont {Diddams},
  \citenamefont {Kitching}, \citenamefont {Papp},\ and\ \citenamefont
  {Hummon}}]{Newman:19}%
  \BibitemOpen
  \bibfield  {author} {\bibinfo {author} {\bibfnamefont {Z.~L.}\ \bibnamefont
  {Newman}}, \bibinfo {author} {\bibfnamefont {V.}~\bibnamefont {Maurice}},
  \bibinfo {author} {\bibfnamefont {T.}~\bibnamefont {Drake}}, \bibinfo
  {author} {\bibfnamefont {J.~R.}\ \bibnamefont {Stone}}, \bibinfo {author}
  {\bibfnamefont {T.~C.}\ \bibnamefont {Briles}}, \bibinfo {author}
  {\bibfnamefont {D.~T.}\ \bibnamefont {Spencer}}, \bibinfo {author}
  {\bibfnamefont {C.}~\bibnamefont {Fredrick}}, \bibinfo {author}
  {\bibfnamefont {Q.}~\bibnamefont {Li}}, \bibinfo {author} {\bibfnamefont
  {D.}~\bibnamefont {Westly}}, \bibinfo {author} {\bibfnamefont {B.~R.}\
  \bibnamefont {Ilic}}, \bibinfo {author} {\bibfnamefont {B.}~\bibnamefont
  {Shen}}, \bibinfo {author} {\bibfnamefont {M.-G.}\ \bibnamefont {Suh}},
  \bibinfo {author} {\bibfnamefont {K.~Y.}\ \bibnamefont {Yang}}, \bibinfo
  {author} {\bibfnamefont {C.}~\bibnamefont {Johnson}}, \bibinfo {author}
  {\bibfnamefont {D.~M.~S.}\ \bibnamefont {Johnson}}, \bibinfo {author}
  {\bibfnamefont {L.}~\bibnamefont {Hollberg}}, \bibinfo {author}
  {\bibfnamefont {K.~J.}\ \bibnamefont {Vahala}}, \bibinfo {author}
  {\bibfnamefont {K.}~\bibnamefont {Srinivasan}}, \bibinfo {author}
  {\bibfnamefont {S.~A.}\ \bibnamefont {Diddams}}, \bibinfo {author}
  {\bibfnamefont {J.}~\bibnamefont {Kitching}}, \bibinfo {author}
  {\bibfnamefont {S.~B.}\ \bibnamefont {Papp}},\ and\ \bibinfo {author}
  {\bibfnamefont {M.~T.}\ \bibnamefont {Hummon}},\ }\bibfield  {title}
  {\bibinfo {title} {Architecture for the photonic integration of an optical
  atomic clock},\ }\href {https://doi.org/10.1364/OPTICA.6.000680} {\bibfield
  {journal} {\bibinfo  {journal} {Optica}\ }\textbf {\bibinfo {volume} {6}},\
  \bibinfo {pages} {680} (\bibinfo {year} {2019})}\BibitemShut {NoStop}%
\bibitem [{\citenamefont {Hausser}\ \emph {et~al.}(2025)\citenamefont
  {Hausser}, \citenamefont {Keller}, \citenamefont {Nordmann}, \citenamefont
  {Bhatt}, \citenamefont {Kiethe}, \citenamefont {Liu}, \citenamefont
  {Richter}, \citenamefont {von Boehn}, \citenamefont {Rahm}, \citenamefont
  {Weyers}, \citenamefont {Benkler}, \citenamefont {Lipphardt}, \citenamefont
  {D\"orscher}, \citenamefont {Stahl}, \citenamefont {Klose}, \citenamefont
  {Lisdat}, \citenamefont {Filzinger}, \citenamefont {Huntemann}, \citenamefont
  {Peik},\ and\ \citenamefont {Mehlst\"aubler}}]{PhysRevLett.134.023201}%
  \BibitemOpen
  \bibfield  {author} {\bibinfo {author} {\bibfnamefont {H.~N.}\ \bibnamefont
  {Hausser}}, \bibinfo {author} {\bibfnamefont {J.}~\bibnamefont {Keller}},
  \bibinfo {author} {\bibfnamefont {T.}~\bibnamefont {Nordmann}}, \bibinfo
  {author} {\bibfnamefont {N.~M.}\ \bibnamefont {Bhatt}}, \bibinfo {author}
  {\bibfnamefont {J.}~\bibnamefont {Kiethe}}, \bibinfo {author} {\bibfnamefont
  {H.}~\bibnamefont {Liu}}, \bibinfo {author} {\bibfnamefont {I.~M.}\
  \bibnamefont {Richter}}, \bibinfo {author} {\bibfnamefont {M.}~\bibnamefont
  {von Boehn}}, \bibinfo {author} {\bibfnamefont {J.}~\bibnamefont {Rahm}},
  \bibinfo {author} {\bibfnamefont {S.}~\bibnamefont {Weyers}}, \bibinfo
  {author} {\bibfnamefont {E.}~\bibnamefont {Benkler}}, \bibinfo {author}
  {\bibfnamefont {B.}~\bibnamefont {Lipphardt}}, \bibinfo {author}
  {\bibfnamefont {S.}~\bibnamefont {D\"orscher}}, \bibinfo {author}
  {\bibfnamefont {K.}~\bibnamefont {Stahl}}, \bibinfo {author} {\bibfnamefont
  {J.}~\bibnamefont {Klose}}, \bibinfo {author} {\bibfnamefont
  {C.}~\bibnamefont {Lisdat}}, \bibinfo {author} {\bibfnamefont
  {M.}~\bibnamefont {Filzinger}}, \bibinfo {author} {\bibfnamefont
  {N.}~\bibnamefont {Huntemann}}, \bibinfo {author} {\bibfnamefont
  {E.}~\bibnamefont {Peik}},\ and\ \bibinfo {author} {\bibfnamefont {T.~E.}\
  \bibnamefont {Mehlst\"aubler}},\ }\bibfield  {title} {\bibinfo {title}
  {$^{115}{\mathrm{in}}^{+}\text{\ensuremath{-}}^{172}{\mathrm{yb}}^{+}$
  coulomb crystal clock with
  $2.5\ifmmode\times\else\texttimes\fi{}{10}^{\ensuremath{-}18}$ systematic
  uncertainty},\ }\href {https://doi.org/10.1103/PhysRevLett.134.023201}
  {\bibfield  {journal} {\bibinfo  {journal} {Phys. Rev. Lett.}\ }\textbf
  {\bibinfo {volume} {134}},\ \bibinfo {pages} {023201} (\bibinfo {year}
  {2025})}\BibitemShut {NoStop}%
\bibitem [{\citenamefont {Chou}\ \emph
  {et~al.}(2010{\natexlab{b}})\citenamefont {Chou}, \citenamefont {Hume},
  \citenamefont {Koelemeij}, \citenamefont {Wineland},\ and\ \citenamefont
  {Rosenband}}]{PhysRevLett.104.070802}%
  \BibitemOpen
  \bibfield  {author} {\bibinfo {author} {\bibfnamefont {C.~W.}\ \bibnamefont
  {Chou}}, \bibinfo {author} {\bibfnamefont {D.~B.}\ \bibnamefont {Hume}},
  \bibinfo {author} {\bibfnamefont {J.~C.~J.}\ \bibnamefont {Koelemeij}},
  \bibinfo {author} {\bibfnamefont {D.~J.}\ \bibnamefont {Wineland}},\ and\
  \bibinfo {author} {\bibfnamefont {T.}~\bibnamefont {Rosenband}},\ }\bibfield
  {title} {\bibinfo {title} {Frequency comparison of two high-accuracy
  ${\mathrm{al}}^{+}$ optical clocks},\ }\href
  {https://doi.org/10.1103/PhysRevLett.104.070802} {\bibfield  {journal}
  {\bibinfo  {journal} {Phys. Rev. Lett.}\ }\textbf {\bibinfo {volume} {104}},\
  \bibinfo {pages} {070802} (\bibinfo {year} {2010}{\natexlab{b}})}\BibitemShut
  {NoStop}%
\bibitem [{\citenamefont {Huang}\ \emph {et~al.}(2022)\citenamefont {Huang},
  \citenamefont {Zhang}, \citenamefont {Zeng}, \citenamefont {Hao},
  \citenamefont {Ma}, \citenamefont {Zhang}, \citenamefont {Guan},
  \citenamefont {Chen}, \citenamefont {Wang},\ and\ \citenamefont
  {Gao}}]{PhysRevApplied.17.034041}%
  \BibitemOpen
  \bibfield  {author} {\bibinfo {author} {\bibfnamefont {Y.}~\bibnamefont
  {Huang}}, \bibinfo {author} {\bibfnamefont {B.}~\bibnamefont {Zhang}},
  \bibinfo {author} {\bibfnamefont {M.}~\bibnamefont {Zeng}}, \bibinfo {author}
  {\bibfnamefont {Y.}~\bibnamefont {Hao}}, \bibinfo {author} {\bibfnamefont
  {Z.}~\bibnamefont {Ma}}, \bibinfo {author} {\bibfnamefont {H.}~\bibnamefont
  {Zhang}}, \bibinfo {author} {\bibfnamefont {H.}~\bibnamefont {Guan}},
  \bibinfo {author} {\bibfnamefont {Z.}~\bibnamefont {Chen}}, \bibinfo {author}
  {\bibfnamefont {M.}~\bibnamefont {Wang}},\ and\ \bibinfo {author}
  {\bibfnamefont {K.}~\bibnamefont {Gao}},\ }\bibfield  {title} {\bibinfo
  {title} {Liquid-nitrogen-cooled $\mathrm{Ca}$${}^{+}$ optical clock with
  systematic uncertainty of
  $3\ifmmode\times\else\texttimes\fi{}{10}^{\ensuremath{-}18}$},\ }\href
  {https://doi.org/10.1103/PhysRevApplied.17.034041} {\bibfield  {journal}
  {\bibinfo  {journal} {Phys. Rev. Appl.}\ }\textbf {\bibinfo {volume} {17}},\
  \bibinfo {pages} {034041} (\bibinfo {year} {2022})}\BibitemShut {NoStop}%
\bibitem [{\citenamefont {Martin}\ \emph {et~al.}(2018)\citenamefont {Martin},
  \citenamefont {Phelps}, \citenamefont {Lemke}, \citenamefont {Bigelow},
  \citenamefont {Stuhl}, \citenamefont {Wojcik}, \citenamefont {Holt},
  \citenamefont {Coddington}, \citenamefont {Bishop},\ and\ \citenamefont
  {Burke}}]{PhysRevApplied.9.014019}%
  \BibitemOpen
  \bibfield  {author} {\bibinfo {author} {\bibfnamefont {K.~W.}\ \bibnamefont
  {Martin}}, \bibinfo {author} {\bibfnamefont {G.}~\bibnamefont {Phelps}},
  \bibinfo {author} {\bibfnamefont {N.~D.}\ \bibnamefont {Lemke}}, \bibinfo
  {author} {\bibfnamefont {M.~S.}\ \bibnamefont {Bigelow}}, \bibinfo {author}
  {\bibfnamefont {B.}~\bibnamefont {Stuhl}}, \bibinfo {author} {\bibfnamefont
  {M.}~\bibnamefont {Wojcik}}, \bibinfo {author} {\bibfnamefont
  {M.}~\bibnamefont {Holt}}, \bibinfo {author} {\bibfnamefont {I.}~\bibnamefont
  {Coddington}}, \bibinfo {author} {\bibfnamefont {M.~W.}\ \bibnamefont
  {Bishop}},\ and\ \bibinfo {author} {\bibfnamefont {J.~H.}\ \bibnamefont
  {Burke}},\ }\bibfield  {title} {\bibinfo {title} {Compact optical atomic
  clock based on a two-photon transition in rubidium},\ }\href
  {https://doi.org/10.1103/PhysRevApplied.9.014019} {\bibfield  {journal}
  {\bibinfo  {journal} {Phys. Rev. Appl.}\ }\textbf {\bibinfo {volume} {9}},\
  \bibinfo {pages} {014019} (\bibinfo {year} {2018})}\BibitemShut {NoStop}%
\bibitem [{\citenamefont {Shi}\ \emph {et~al.}(2024)\citenamefont {Shi},
  \citenamefont {Wei}, \citenamefont {Qin}, \citenamefont {Liu}, \citenamefont
  {Chen}, \citenamefont {Cao}, \citenamefont {Shi}, \citenamefont {Liu},\ and\
  \citenamefont {Chen}}]{Shi:24}%
  \BibitemOpen
  \bibfield  {author} {\bibinfo {author} {\bibfnamefont {T.}~\bibnamefont
  {Shi}}, \bibinfo {author} {\bibfnamefont {Q.}~\bibnamefont {Wei}}, \bibinfo
  {author} {\bibfnamefont {X.}~\bibnamefont {Qin}}, \bibinfo {author}
  {\bibfnamefont {Z.}~\bibnamefont {Liu}}, \bibinfo {author} {\bibfnamefont
  {K.}~\bibnamefont {Chen}}, \bibinfo {author} {\bibfnamefont {S.}~\bibnamefont
  {Cao}}, \bibinfo {author} {\bibfnamefont {H.}~\bibnamefont {Shi}}, \bibinfo
  {author} {\bibfnamefont {Z.}~\bibnamefont {Liu}},\ and\ \bibinfo {author}
  {\bibfnamefont {J.}~\bibnamefont {Chen}},\ }\bibfield  {title} {\bibinfo
  {title} {Dual-frequency optical-microwave atomic clocks based on cesium
  atoms},\ }\href {https://doi.org/10.1364/PRJ.528942} {\bibfield  {journal}
  {\bibinfo  {journal} {Photon. Res.}\ }\textbf {\bibinfo {volume} {12}},\
  \bibinfo {pages} {1972} (\bibinfo {year} {2024})}\BibitemShut {NoStop}%
\bibitem [{\citenamefont {Hao}\ \emph {et~al.}(2024)\citenamefont {Hao},
  \citenamefont {Yang}, \citenamefont {Ruan}, \citenamefont {Yun},\ and\
  \citenamefont {Zhang}}]{PhysRevApplied.21.024003}%
  \BibitemOpen
  \bibfield  {author} {\bibinfo {author} {\bibfnamefont {Q.}~\bibnamefont
  {Hao}}, \bibinfo {author} {\bibfnamefont {S.}~\bibnamefont {Yang}}, \bibinfo
  {author} {\bibfnamefont {J.}~\bibnamefont {Ruan}}, \bibinfo {author}
  {\bibfnamefont {P.}~\bibnamefont {Yun}},\ and\ \bibinfo {author}
  {\bibfnamefont {S.}~\bibnamefont {Zhang}},\ }\bibfield  {title} {\bibinfo
  {title} {Integrated pulsed optically pumped rb atomic clock with frequency
  stability of ${10}^{\ensuremath{-}15}$},\ }\href
  {https://doi.org/10.1103/PhysRevApplied.21.024003} {\bibfield  {journal}
  {\bibinfo  {journal} {Phys. Rev. Appl.}\ }\textbf {\bibinfo {volume} {21}},\
  \bibinfo {pages} {024003} (\bibinfo {year} {2024})}\BibitemShut {NoStop}%
\bibitem [{\citenamefont {Phelps}\ \emph {et~al.}(2018)\citenamefont {Phelps},
  \citenamefont {Lemke}, \citenamefont {Erickson}, \citenamefont {Burke},\ and\
  \citenamefont {Martin}}]{https://doi.org/10.1002/navi.215}%
  \BibitemOpen
  \bibfield  {author} {\bibinfo {author} {\bibfnamefont {G.}~\bibnamefont
  {Phelps}}, \bibinfo {author} {\bibfnamefont {N.}~\bibnamefont {Lemke}},
  \bibinfo {author} {\bibfnamefont {C.}~\bibnamefont {Erickson}}, \bibinfo
  {author} {\bibfnamefont {J.}~\bibnamefont {Burke}},\ and\ \bibinfo {author}
  {\bibfnamefont {K.}~\bibnamefont {Martin}},\ }\bibfield  {title} {\bibinfo
  {title} {Compact optical clock with $5\times 10^{-13}$ instability at 1 s},\
  }\href {https://doi.org/https://doi.org/10.1002/navi.215} {\bibfield
  {journal} {\bibinfo  {journal} {NAVIGATION}\ }\textbf {\bibinfo {volume}
  {65}},\ \bibinfo {pages} {49} (\bibinfo {year} {2018})}\BibitemShut {NoStop}%
\bibitem [{\citenamefont {Newman}\ \emph {et~al.}(2021)\citenamefont {Newman},
  \citenamefont {Maurice}, \citenamefont {Fredrick}, \citenamefont {Fortier},
  \citenamefont {Leopardi}, \citenamefont {Hollberg}, \citenamefont {Diddams},
  \citenamefont {Kitching},\ and\ \citenamefont {Hummon}}]{Newman:21}%
  \BibitemOpen
  \bibfield  {author} {\bibinfo {author} {\bibfnamefont {Z.~L.}\ \bibnamefont
  {Newman}}, \bibinfo {author} {\bibfnamefont {V.}~\bibnamefont {Maurice}},
  \bibinfo {author} {\bibfnamefont {C.}~\bibnamefont {Fredrick}}, \bibinfo
  {author} {\bibfnamefont {T.}~\bibnamefont {Fortier}}, \bibinfo {author}
  {\bibfnamefont {H.}~\bibnamefont {Leopardi}}, \bibinfo {author}
  {\bibfnamefont {L.}~\bibnamefont {Hollberg}}, \bibinfo {author}
  {\bibfnamefont {S.~A.}\ \bibnamefont {Diddams}}, \bibinfo {author}
  {\bibfnamefont {J.}~\bibnamefont {Kitching}},\ and\ \bibinfo {author}
  {\bibfnamefont {M.~T.}\ \bibnamefont {Hummon}},\ }\bibfield  {title}
  {\bibinfo {title} {High-performance, compact optical standard},\ }\href
  {https://doi.org/10.1364/OL.435603} {\bibfield  {journal} {\bibinfo
  {journal} {Opt. Lett.}\ }\textbf {\bibinfo {volume} {46}},\ \bibinfo {pages}
  {4702} (\bibinfo {year} {2021})}\BibitemShut {NoStop}%
\bibitem [{\citenamefont {Callejo}\ \emph {et~al.}(2025)\citenamefont
  {Callejo}, \citenamefont {Mursa}, \citenamefont {Vicarini}, \citenamefont
  {Klinger}, \citenamefont {Tanguy}, \citenamefont {Millo}, \citenamefont
  {Passilly},\ and\ \citenamefont {Boudot}}]{Callejo:25}%
  \BibitemOpen
  \bibfield  {author} {\bibinfo {author} {\bibfnamefont {M.}~\bibnamefont
  {Callejo}}, \bibinfo {author} {\bibfnamefont {A.}~\bibnamefont {Mursa}},
  \bibinfo {author} {\bibfnamefont {R.}~\bibnamefont {Vicarini}}, \bibinfo
  {author} {\bibfnamefont {E.}~\bibnamefont {Klinger}}, \bibinfo {author}
  {\bibfnamefont {Q.}~\bibnamefont {Tanguy}}, \bibinfo {author} {\bibfnamefont
  {J.}~\bibnamefont {Millo}}, \bibinfo {author} {\bibfnamefont
  {N.}~\bibnamefont {Passilly}},\ and\ \bibinfo {author} {\bibfnamefont
  {R.}~\bibnamefont {Boudot}},\ }\bibfield  {title} {\bibinfo {title}
  {Short-term stability of a microcell optical reference based on the rb atom
  two-photon transition at 778 nm},\ }\href
  {https://doi.org/10.1364/JOSAB.533904} {\bibfield  {journal} {\bibinfo
  {journal} {J. Opt. Soc. Am. B}\ }\textbf {\bibinfo {volume} {42}},\ \bibinfo
  {pages} {151} (\bibinfo {year} {2025})}\BibitemShut {NoStop}%
\bibitem [{\citenamefont {Martin}\ \emph {et~al.}(2019)\citenamefont {Martin},
  \citenamefont {Stuhl}, \citenamefont {Eugenio}, \citenamefont {Safronova},
  \citenamefont {Phelps}, \citenamefont {Burke},\ and\ \citenamefont
  {Lemke}}]{PhysRevA.100.023417}%
  \BibitemOpen
  \bibfield  {author} {\bibinfo {author} {\bibfnamefont {K.~W.}\ \bibnamefont
  {Martin}}, \bibinfo {author} {\bibfnamefont {B.}~\bibnamefont {Stuhl}},
  \bibinfo {author} {\bibfnamefont {J.}~\bibnamefont {Eugenio}}, \bibinfo
  {author} {\bibfnamefont {M.~S.}\ \bibnamefont {Safronova}}, \bibinfo {author}
  {\bibfnamefont {G.}~\bibnamefont {Phelps}}, \bibinfo {author} {\bibfnamefont
  {J.~H.}\ \bibnamefont {Burke}},\ and\ \bibinfo {author} {\bibfnamefont
  {N.~D.}\ \bibnamefont {Lemke}},\ }\bibfield  {title} {\bibinfo {title}
  {Frequency shifts due to stark effects on a rubidium two-photon transition},\
  }\href {https://doi.org/10.1103/PhysRevA.100.023417} {\bibfield  {journal}
  {\bibinfo  {journal} {Phys. Rev. A}\ }\textbf {\bibinfo {volume} {100}},\
  \bibinfo {pages} {023417} (\bibinfo {year} {2019})}\BibitemShut {NoStop}%
\bibitem [{\citenamefont {Yudin}\ \emph {et~al.}(2020)\citenamefont {Yudin},
  \citenamefont {Basalaev}, \citenamefont {Taichenachev}, \citenamefont
  {Pollock}, \citenamefont {Newman}, \citenamefont {Shuker}, \citenamefont
  {Hansen}, \citenamefont {Hummon}, \citenamefont {Boudot}, \citenamefont
  {Donley},\ and\ \citenamefont {Kitching}}]{PhysRevApplied.14.024001}%
  \BibitemOpen
  \bibfield  {author} {\bibinfo {author} {\bibfnamefont {V.}~\bibnamefont
  {Yudin}}, \bibinfo {author} {\bibfnamefont {M.~Y.}\ \bibnamefont {Basalaev}},
  \bibinfo {author} {\bibfnamefont {A.}~\bibnamefont {Taichenachev}}, \bibinfo
  {author} {\bibfnamefont {J.}~\bibnamefont {Pollock}}, \bibinfo {author}
  {\bibfnamefont {Z.}~\bibnamefont {Newman}}, \bibinfo {author} {\bibfnamefont
  {M.}~\bibnamefont {Shuker}}, \bibinfo {author} {\bibfnamefont
  {A.}~\bibnamefont {Hansen}}, \bibinfo {author} {\bibfnamefont
  {M.}~\bibnamefont {Hummon}}, \bibinfo {author} {\bibfnamefont
  {R.}~\bibnamefont {Boudot}}, \bibinfo {author} {\bibfnamefont
  {E.}~\bibnamefont {Donley}},\ and\ \bibinfo {author} {\bibfnamefont
  {J.}~\bibnamefont {Kitching}},\ }\bibfield  {title} {\bibinfo {title}
  {General methods for suppressing the light shift in atomic clocks using power
  modulation},\ }\href {https://doi.org/10.1103/PhysRevApplied.14.024001}
  {\bibfield  {journal} {\bibinfo  {journal} {Phys. Rev. Appl.}\ }\textbf
  {\bibinfo {volume} {14}},\ \bibinfo {pages} {024001} (\bibinfo {year}
  {2020})}\BibitemShut {NoStop}%
\bibitem [{\citenamefont {Gerginov}\ and\ \citenamefont
  {Beloy}(2018)}]{PhysRevApplied.10.014031}%
  \BibitemOpen
  \bibfield  {author} {\bibinfo {author} {\bibfnamefont {V.}~\bibnamefont
  {Gerginov}}\ and\ \bibinfo {author} {\bibfnamefont {K.}~\bibnamefont
  {Beloy}},\ }\bibfield  {title} {\bibinfo {title} {Two-photon optical
  frequency reference with active ac stark shift cancellation},\ }\href
  {https://doi.org/10.1103/PhysRevApplied.10.014031} {\bibfield  {journal}
  {\bibinfo  {journal} {Phys. Rev. Appl.}\ }\textbf {\bibinfo {volume} {10}},\
  \bibinfo {pages} {014031} (\bibinfo {year} {2018})}\BibitemShut {NoStop}%
\bibitem [{\citenamefont {Hafiz}\ \emph {et~al.}(2016)\citenamefont {Hafiz},
  \citenamefont {Coget}, \citenamefont {Clercq},\ and\ \citenamefont
  {Boudot}}]{Hafiz:16}%
  \BibitemOpen
  \bibfield  {author} {\bibinfo {author} {\bibfnamefont {M.~A.}\ \bibnamefont
  {Hafiz}}, \bibinfo {author} {\bibfnamefont {G.}~\bibnamefont {Coget}},
  \bibinfo {author} {\bibfnamefont {E.~D.}\ \bibnamefont {Clercq}},\ and\
  \bibinfo {author} {\bibfnamefont {R.}~\bibnamefont {Boudot}},\ }\bibfield
  {title} {\bibinfo {title} {Doppler-free spectroscopy on the cs d1 line with a
  dual-frequency laser},\ }\href {https://doi.org/10.1364/OL.41.002982}
  {\bibfield  {journal} {\bibinfo  {journal} {Opt. Lett.}\ }\textbf {\bibinfo
  {volume} {41}},\ \bibinfo {pages} {2982} (\bibinfo {year}
  {2016})}\BibitemShut {NoStop}%
\bibitem [{\citenamefont {Lemke}\ \emph {et~al.}(2022)\citenamefont {Lemke},
  \citenamefont {Martin}, \citenamefont {Beard}, \citenamefont {Stuhl},
  \citenamefont {Metcalf},\ and\ \citenamefont {Elgin}}]{s22051982}%
  \BibitemOpen
  \bibfield  {author} {\bibinfo {author} {\bibfnamefont {N.~D.}\ \bibnamefont
  {Lemke}}, \bibinfo {author} {\bibfnamefont {K.~W.}\ \bibnamefont {Martin}},
  \bibinfo {author} {\bibfnamefont {R.}~\bibnamefont {Beard}}, \bibinfo
  {author} {\bibfnamefont {B.~K.}\ \bibnamefont {Stuhl}}, \bibinfo {author}
  {\bibfnamefont {A.~J.}\ \bibnamefont {Metcalf}},\ and\ \bibinfo {author}
  {\bibfnamefont {J.~D.}\ \bibnamefont {Elgin}},\ }\bibfield  {title} {\bibinfo
  {title} {Measurement of optical rubidium clock frequency spanning 65 days},\
  }\bibfield  {journal} {\bibinfo  {journal} {Sensors}\ }\textbf {\bibinfo
  {volume} {22}},\ \href {https://doi.org/10.3390/s22051982}
  {10.3390/s22051982} (\bibinfo {year} {2022})\BibitemShut {NoStop}%
\bibitem [{\citenamefont {Yudin}\ \emph {et~al.}(2023)\citenamefont {Yudin},
  \citenamefont {Basalaev}, \citenamefont {Taichenachev}, \citenamefont
  {Prudnikov}, \citenamefont {Radnatarov}, \citenamefont {Kobtsev},
  \citenamefont {Ignatovich},\ and\ \citenamefont
  {Skvortsov}}]{PhysRevA.108.013103}%
  \BibitemOpen
  \bibfield  {author} {\bibinfo {author} {\bibfnamefont {V.~I.}\ \bibnamefont
  {Yudin}}, \bibinfo {author} {\bibfnamefont {M.~Y.}\ \bibnamefont {Basalaev}},
  \bibinfo {author} {\bibfnamefont {A.~V.}\ \bibnamefont {Taichenachev}},
  \bibinfo {author} {\bibfnamefont {O.~N.}\ \bibnamefont {Prudnikov}}, \bibinfo
  {author} {\bibfnamefont {D.~A.}\ \bibnamefont {Radnatarov}}, \bibinfo
  {author} {\bibfnamefont {S.~M.}\ \bibnamefont {Kobtsev}}, \bibinfo {author}
  {\bibfnamefont {S.~M.}\ \bibnamefont {Ignatovich}},\ and\ \bibinfo {author}
  {\bibfnamefont {M.~N.}\ \bibnamefont {Skvortsov}},\ }\bibfield  {title}
  {\bibinfo {title} {Frequency shift caused by the line-shape asymmetry of the
  resonance of coherent population trapping},\ }\href
  {https://doi.org/10.1103/PhysRevA.108.013103} {\bibfield  {journal} {\bibinfo
   {journal} {Phys. Rev. A}\ }\textbf {\bibinfo {volume} {108}},\ \bibinfo
  {pages} {013103} (\bibinfo {year} {2023})}\BibitemShut {NoStop}%
\bibitem [{\citenamefont {Abdel~Hafiz}\ and\ \citenamefont
  {Boudot}(2015)}]{10.1063/1.4931768}%
  \BibitemOpen
  \bibfield  {author} {\bibinfo {author} {\bibfnamefont {M.}~\bibnamefont
  {Abdel~Hafiz}}\ and\ \bibinfo {author} {\bibfnamefont {R.}~\bibnamefont
  {Boudot}},\ }\bibfield  {title} {\bibinfo {title} {{A coherent population
  trapping Cs vapor cell atomic clock based on push-pull optical pumping}},\
  }\href {https://doi.org/10.1063/1.4931768} {\bibfield  {journal} {\bibinfo
  {journal} {Journal of Applied Physics}\ }\textbf {\bibinfo {volume} {118}},\
  \bibinfo {pages} {124903} (\bibinfo {year} {2015})}\BibitemShut {NoStop}%
\bibitem [{\citenamefont {Yudin}\ \emph {et~al.}(2017)\citenamefont {Yudin},
  \citenamefont {Taichenachev}, \citenamefont {Basalaev},\ and\ \citenamefont
  {Kovalenko}}]{Yudin:17}%
  \BibitemOpen
  \bibfield  {author} {\bibinfo {author} {\bibfnamefont {V.~I.}\ \bibnamefont
  {Yudin}}, \bibinfo {author} {\bibfnamefont {A.~V.}\ \bibnamefont
  {Taichenachev}}, \bibinfo {author} {\bibfnamefont {M.~Y.}\ \bibnamefont
  {Basalaev}},\ and\ \bibinfo {author} {\bibfnamefont {D.~V.}\ \bibnamefont
  {Kovalenko}},\ }\bibfield  {title} {\bibinfo {title} {Dynamic regime of
  coherent population trapping and optimization of frequency modulation
  parameters in atomic clocks},\ }\href {https://doi.org/10.1364/OE.25.002742}
  {\bibfield  {journal} {\bibinfo  {journal} {Opt. Express}\ }\textbf {\bibinfo
  {volume} {25}},\ \bibinfo {pages} {2742} (\bibinfo {year}
  {2017})}\BibitemShut {NoStop}%
\bibitem [{\citenamefont {Fang}\ \emph {et~al.}(2021)\citenamefont {Fang},
  \citenamefont {Han}, \citenamefont {Jiang}, \citenamefont {Qiu},
  \citenamefont {Guo}, \citenamefont {Zhao}, \citenamefont {Huang},
  \citenamefont {Lu},\ and\ \citenamefont {Lee}}]{Fang2021}%
  \BibitemOpen
  \bibfield  {author} {\bibinfo {author} {\bibfnamefont {R.}~\bibnamefont
  {Fang}}, \bibinfo {author} {\bibfnamefont {C.}~\bibnamefont {Han}}, \bibinfo
  {author} {\bibfnamefont {X.}~\bibnamefont {Jiang}}, \bibinfo {author}
  {\bibfnamefont {Y.}~\bibnamefont {Qiu}}, \bibinfo {author} {\bibfnamefont
  {Y.}~\bibnamefont {Guo}}, \bibinfo {author} {\bibfnamefont {M.}~\bibnamefont
  {Zhao}}, \bibinfo {author} {\bibfnamefont {J.}~\bibnamefont {Huang}},
  \bibinfo {author} {\bibfnamefont {B.}~\bibnamefont {Lu}},\ and\ \bibinfo
  {author} {\bibfnamefont {C.}~\bibnamefont {Lee}},\ }\bibfield  {title}
  {\bibinfo {title} {Temporal analog of fabry-p{\'e}rot resonator via coherent
  population trapping},\ }\href {https://doi.org/10.1038/s41534-021-00479-y}
  {\bibfield  {journal} {\bibinfo  {journal} {npj Quantum Information}\
  }\textbf {\bibinfo {volume} {7}},\ \bibinfo {pages} {143} (\bibinfo {year}
  {2021})}\BibitemShut {NoStop}%
\bibitem [{\citenamefont {Huelga}\ \emph {et~al.}(1997)\citenamefont {Huelga},
  \citenamefont {Macchiavello}, \citenamefont {Pellizzari}, \citenamefont
  {Ekert}, \citenamefont {Plenio},\ and\ \citenamefont
  {Cirac}}]{PhysRevLett.79.3865}%
  \BibitemOpen
  \bibfield  {author} {\bibinfo {author} {\bibfnamefont {S.~F.}\ \bibnamefont
  {Huelga}}, \bibinfo {author} {\bibfnamefont {C.}~\bibnamefont
  {Macchiavello}}, \bibinfo {author} {\bibfnamefont {T.}~\bibnamefont
  {Pellizzari}}, \bibinfo {author} {\bibfnamefont {A.~K.}\ \bibnamefont
  {Ekert}}, \bibinfo {author} {\bibfnamefont {M.~B.}\ \bibnamefont {Plenio}},\
  and\ \bibinfo {author} {\bibfnamefont {J.~I.}\ \bibnamefont {Cirac}},\
  }\bibfield  {title} {\bibinfo {title} {Improvement of frequency standards
  with quantum entanglement},\ }\href
  {https://doi.org/10.1103/PhysRevLett.79.3865} {\bibfield  {journal} {\bibinfo
   {journal} {Phys. Rev. Lett.}\ }\textbf {\bibinfo {volume} {79}},\ \bibinfo
  {pages} {3865} (\bibinfo {year} {1997})}\BibitemShut {NoStop}%
\bibitem [{\citenamefont {Perrella}\ \emph {et~al.}(2019)\citenamefont
  {Perrella}, \citenamefont {Light}, \citenamefont {Anstie}, \citenamefont
  {Baynes}, \citenamefont {White},\ and\ \citenamefont
  {Luiten}}]{PhysRevApplied.12.054063}%
  \BibitemOpen
  \bibfield  {author} {\bibinfo {author} {\bibfnamefont {C.}~\bibnamefont
  {Perrella}}, \bibinfo {author} {\bibfnamefont {P.}~\bibnamefont {Light}},
  \bibinfo {author} {\bibfnamefont {J.}~\bibnamefont {Anstie}}, \bibinfo
  {author} {\bibfnamefont {F.}~\bibnamefont {Baynes}}, \bibinfo {author}
  {\bibfnamefont {R.}~\bibnamefont {White}},\ and\ \bibinfo {author}
  {\bibfnamefont {A.}~\bibnamefont {Luiten}},\ }\bibfield  {title} {\bibinfo
  {title} {Dichroic two-photon rubidium frequency standard},\ }\href
  {https://doi.org/10.1103/PhysRevApplied.12.054063} {\bibfield  {journal}
  {\bibinfo  {journal} {Phys. Rev. Appl.}\ }\textbf {\bibinfo {volume} {12}},\
  \bibinfo {pages} {054063} (\bibinfo {year} {2019})}\BibitemShut {NoStop}%
\bibitem [{\citenamefont {Holzwarth}\ \emph {et~al.}(2001)\citenamefont
  {Holzwarth}, \citenamefont {Zimmermann}, \citenamefont {Udem},\ and\
  \citenamefont {Hansch}}]{970894}%
  \BibitemOpen
  \bibfield  {author} {\bibinfo {author} {\bibfnamefont {R.}~\bibnamefont
  {Holzwarth}}, \bibinfo {author} {\bibfnamefont {M.}~\bibnamefont
  {Zimmermann}}, \bibinfo {author} {\bibfnamefont {T.}~\bibnamefont {Udem}},\
  and\ \bibinfo {author} {\bibfnamefont {T.}~\bibnamefont {Hansch}},\
  }\bibfield  {title} {\bibinfo {title} {Optical clockworks and the measurement
  of laser frequencies with a mode-locked frequency comb},\ }\href
  {https://doi.org/10.1109/3.970894} {\bibfield  {journal} {\bibinfo  {journal}
  {IEEE Journal of Quantum Electronics}\ }\textbf {\bibinfo {volume} {37}},\
  \bibinfo {pages} {1493} (\bibinfo {year} {2001})}\BibitemShut {NoStop}%
\bibitem [{\citenamefont {Fortier}\ and\ \citenamefont
  {Baumann}(2019)}]{Fortier2019}%
  \BibitemOpen
  \bibfield  {author} {\bibinfo {author} {\bibfnamefont {T.}~\bibnamefont
  {Fortier}}\ and\ \bibinfo {author} {\bibfnamefont {E.}~\bibnamefont
  {Baumann}},\ }\bibfield  {title} {\bibinfo {title} {20 years of developments
  in optical frequency comb technology and applications},\ }\href
  {https://doi.org/10.1038/s42005-019-0249-y} {\bibfield  {journal} {\bibinfo
  {journal} {Communications Physics}\ }\textbf {\bibinfo {volume} {2}},\
  \bibinfo {pages} {153} (\bibinfo {year} {2019})}\BibitemShut {NoStop}%
\bibitem [{\citenamefont {Phillips}\ \emph {et~al.}(2011)\citenamefont
  {Phillips}, \citenamefont {Gorshkov},\ and\ \citenamefont
  {Novikova}}]{PhysRevA.83.063823}%
  \BibitemOpen
  \bibfield  {author} {\bibinfo {author} {\bibfnamefont {N.~B.}\ \bibnamefont
  {Phillips}}, \bibinfo {author} {\bibfnamefont {A.~V.}\ \bibnamefont
  {Gorshkov}},\ and\ \bibinfo {author} {\bibfnamefont {I.}~\bibnamefont
  {Novikova}},\ }\bibfield  {title} {\bibinfo {title} {Light storage in an
  optically thick atomic ensemble under conditions of electromagnetically
  induced transparency and four-wave mixing},\ }\href
  {https://doi.org/10.1103/PhysRevA.83.063823} {\bibfield  {journal} {\bibinfo
  {journal} {Phys. Rev. A}\ }\textbf {\bibinfo {volume} {83}},\ \bibinfo
  {pages} {063823} (\bibinfo {year} {2011})}\BibitemShut {NoStop}%
\bibitem [{\citenamefont {Abram}(1987)}]{PhysRevA.35.4661}%
  \BibitemOpen
  \bibfield  {author} {\bibinfo {author} {\bibfnamefont {I.}~\bibnamefont
  {Abram}},\ }\bibfield  {title} {\bibinfo {title} {Quantum theory of light
  propagation: Linear medium},\ }\href
  {https://doi.org/10.1103/PhysRevA.35.4661} {\bibfield  {journal} {\bibinfo
  {journal} {Phys. Rev. A}\ }\textbf {\bibinfo {volume} {35}},\ \bibinfo
  {pages} {4661} (\bibinfo {year} {1987})}\BibitemShut {NoStop}%
\bibitem [{\citenamefont {Blow}\ \emph {et~al.}(1990)\citenamefont {Blow},
  \citenamefont {Loudon}, \citenamefont {Phoenix},\ and\ \citenamefont
  {Shepherd}}]{PhysRevA.42.4102}%
  \BibitemOpen
  \bibfield  {author} {\bibinfo {author} {\bibfnamefont {K.~J.}\ \bibnamefont
  {Blow}}, \bibinfo {author} {\bibfnamefont {R.}~\bibnamefont {Loudon}},
  \bibinfo {author} {\bibfnamefont {S.~J.~D.}\ \bibnamefont {Phoenix}},\ and\
  \bibinfo {author} {\bibfnamefont {T.~J.}\ \bibnamefont {Shepherd}},\
  }\bibfield  {title} {\bibinfo {title} {Continuum fields in quantum optics},\
  }\href {https://doi.org/10.1103/PhysRevA.42.4102} {\bibfield  {journal}
  {\bibinfo  {journal} {Phys. Rev. A}\ }\textbf {\bibinfo {volume} {42}},\
  \bibinfo {pages} {4102} (\bibinfo {year} {1990})}\BibitemShut {NoStop}%
\bibitem [{\citenamefont {Drummond}\ and\ \citenamefont
  {Carter}(1987)}]{Drummond:87}%
  \BibitemOpen
  \bibfield  {author} {\bibinfo {author} {\bibfnamefont {P.~D.}\ \bibnamefont
  {Drummond}}\ and\ \bibinfo {author} {\bibfnamefont {S.~J.}\ \bibnamefont
  {Carter}},\ }\bibfield  {title} {\bibinfo {title} {Quantum-field theory of
  squeezing in solitons},\ }\href {https://doi.org/10.1364/JOSAB.4.001565}
  {\bibfield  {journal} {\bibinfo  {journal} {J. Opt. Soc. Am. B}\ }\textbf
  {\bibinfo {volume} {4}},\ \bibinfo {pages} {1565} (\bibinfo {year}
  {1987})}\BibitemShut {NoStop}%
\bibitem [{\citenamefont {Fleischhauer}\ and\ \citenamefont
  {Scully}(1994)}]{PhysRevA.49.1973}%
  \BibitemOpen
  \bibfield  {author} {\bibinfo {author} {\bibfnamefont {M.}~\bibnamefont
  {Fleischhauer}}\ and\ \bibinfo {author} {\bibfnamefont {M.~O.}\ \bibnamefont
  {Scully}},\ }\bibfield  {title} {\bibinfo {title} {Quantum sensitivity limits
  of an optical magnetometer based on atomic phase coherence},\ }\href
  {https://doi.org/10.1103/PhysRevA.49.1973} {\bibfield  {journal} {\bibinfo
  {journal} {Phys. Rev. A}\ }\textbf {\bibinfo {volume} {49}},\ \bibinfo
  {pages} {1973} (\bibinfo {year} {1994})}\BibitemShut {NoStop}%
\bibitem [{\citenamefont {Gardiner}\ and\ \citenamefont
  {Zoller}(2004)}]{gardiner2004quantum}%
  \BibitemOpen
  \bibfield  {author} {\bibinfo {author} {\bibfnamefont {C.}~\bibnamefont
  {Gardiner}}\ and\ \bibinfo {author} {\bibfnamefont {P.}~\bibnamefont
  {Zoller}},\ }\href@noop {} {\emph {\bibinfo {title} {Quantum noise: a
  handbook of Markovian and non-Markovian quantum stochastic methods with
  applications to quantum optics}}}\ (\bibinfo  {publisher} {Springer Science
  \& Business Media},\ \bibinfo {year} {2004})\BibitemShut {NoStop}%
\bibitem [{\citenamefont {Boller}\ \emph {et~al.}(1991)\citenamefont {Boller},
  \citenamefont {Imamo\ifmmode~\breve{g}\else \u{g}\fi{}lu},\ and\
  \citenamefont {Harris}}]{PhysRevLett.66.2593}%
  \BibitemOpen
  \bibfield  {author} {\bibinfo {author} {\bibfnamefont {K.-J.}\ \bibnamefont
  {Boller}}, \bibinfo {author} {\bibfnamefont {A.}~\bibnamefont
  {Imamo\ifmmode~\breve{g}\else \u{g}\fi{}lu}},\ and\ \bibinfo {author}
  {\bibfnamefont {S.~E.}\ \bibnamefont {Harris}},\ }\bibfield  {title}
  {\bibinfo {title} {Observation of electromagnetically induced transparency},\
  }\href {https://doi.org/10.1103/PhysRevLett.66.2593} {\bibfield  {journal}
  {\bibinfo  {journal} {Phys. Rev. Lett.}\ }\textbf {\bibinfo {volume} {66}},\
  \bibinfo {pages} {2593} (\bibinfo {year} {1991})}\BibitemShut {NoStop}%
\bibitem [{\citenamefont {Drain}\ and\ \citenamefont {Moss}(1972)}]{Drain1972}%
  \BibitemOpen
  \bibfield  {author} {\bibinfo {author} {\bibfnamefont {L.~E.}\ \bibnamefont
  {Drain}}\ and\ \bibinfo {author} {\bibfnamefont {B.~C.}\ \bibnamefont
  {Moss}},\ }\bibfield  {title} {\bibinfo {title} {The frequency shifting of
  laser light by electro-optic techniques},\ }\href
  {https://doi.org/10.1007/BF01414146} {\bibfield  {journal} {\bibinfo
  {journal} {Opto-electronics}\ }\textbf {\bibinfo {volume} {4}},\ \bibinfo
  {pages} {429} (\bibinfo {year} {1972})}\BibitemShut {NoStop}%
\bibitem [{\citenamefont {Zeng}\ \emph {et~al.}(2020)\citenamefont {Zeng},
  \citenamefont {Zhang}, \citenamefont {Zhang}, \citenamefont {Zhang},
  \citenamefont {Zhang}, \citenamefont {Sun},\ and\ \citenamefont
  {Liu}}]{Zeng:20}%
  \BibitemOpen
  \bibfield  {author} {\bibinfo {author} {\bibfnamefont {Z.}~\bibnamefont
  {Zeng}}, \bibinfo {author} {\bibfnamefont {Z.}~\bibnamefont {Zhang}},
  \bibinfo {author} {\bibfnamefont {L.}~\bibnamefont {Zhang}}, \bibinfo
  {author} {\bibfnamefont {S.}~\bibnamefont {Zhang}}, \bibinfo {author}
  {\bibfnamefont {Y.}~\bibnamefont {Zhang}}, \bibinfo {author} {\bibfnamefont
  {B.}~\bibnamefont {Sun}},\ and\ \bibinfo {author} {\bibfnamefont
  {Y.}~\bibnamefont {Liu}},\ }\bibfield  {title} {\bibinfo {title} {Stable and
  finely tunable optoelectronic oscillator based on stimulated brillouin
  scattering and an electro-optic frequency shift},\ }\href
  {https://doi.org/10.1364/AO.378196} {\bibfield  {journal} {\bibinfo
  {journal} {Appl. Opt.}\ }\textbf {\bibinfo {volume} {59}},\ \bibinfo {pages}
  {589} (\bibinfo {year} {2020})}\BibitemShut {NoStop}%
\bibitem [{\citenamefont {Jones}\ \emph {et~al.}(2005)\citenamefont {Jones},
  \citenamefont {Moll}, \citenamefont {Thorpe},\ and\ \citenamefont
  {Ye}}]{PhysRevLett.94.193201}%
  \BibitemOpen
  \bibfield  {author} {\bibinfo {author} {\bibfnamefont {R.~J.}\ \bibnamefont
  {Jones}}, \bibinfo {author} {\bibfnamefont {K.~D.}\ \bibnamefont {Moll}},
  \bibinfo {author} {\bibfnamefont {M.~J.}\ \bibnamefont {Thorpe}},\ and\
  \bibinfo {author} {\bibfnamefont {J.}~\bibnamefont {Ye}},\ }\bibfield
  {title} {\bibinfo {title} {Phase-coherent frequency combs in the vacuum
  ultraviolet via high-harmonic generation inside a femtosecond enhancement
  cavity},\ }\href {https://doi.org/10.1103/PhysRevLett.94.193201} {\bibfield
  {journal} {\bibinfo  {journal} {Phys. Rev. Lett.}\ }\textbf {\bibinfo
  {volume} {94}},\ \bibinfo {pages} {193201} (\bibinfo {year}
  {2005})}\BibitemShut {NoStop}%
\bibitem [{\citenamefont {Schulte}\ \emph {et~al.}(2020)\citenamefont
  {Schulte}, \citenamefont {Lisdat}, \citenamefont {Schmidt}, \citenamefont
  {Sterr},\ and\ \citenamefont {Hammerer}}]{Schulte2020}%
  \BibitemOpen
  \bibfield  {author} {\bibinfo {author} {\bibfnamefont {M.}~\bibnamefont
  {Schulte}}, \bibinfo {author} {\bibfnamefont {C.}~\bibnamefont {Lisdat}},
  \bibinfo {author} {\bibfnamefont {P.~O.}\ \bibnamefont {Schmidt}}, \bibinfo
  {author} {\bibfnamefont {U.}~\bibnamefont {Sterr}},\ and\ \bibinfo {author}
  {\bibfnamefont {K.}~\bibnamefont {Hammerer}},\ }\bibfield  {title} {\bibinfo
  {title} {Prospects and challenges for squeezing-enhanced optical atomic
  clocks},\ }\href {https://doi.org/10.1038/s41467-020-19403-7} {\bibfield
  {journal} {\bibinfo  {journal} {Nature Communications}\ }\textbf {\bibinfo
  {volume} {11}},\ \bibinfo {pages} {5955} (\bibinfo {year}
  {2020})}\BibitemShut {NoStop}%
\bibitem [{\citenamefont {Farkas}\ \emph {et~al.}(2010)\citenamefont {Farkas},
  \citenamefont {Zozulya},\ and\ \citenamefont {Anderson}}]{Farkas2010}%
  \BibitemOpen
  \bibfield  {author} {\bibinfo {author} {\bibfnamefont {D.~M.}\ \bibnamefont
  {Farkas}}, \bibinfo {author} {\bibfnamefont {A.}~\bibnamefont {Zozulya}},\
  and\ \bibinfo {author} {\bibfnamefont {D.~Z.}\ \bibnamefont {Anderson}},\
  }\bibfield  {title} {\bibinfo {title} {A compact microchip atomic clock based
  on all-optical interrogation of ultra-cold trapped rb atoms},\ }\href
  {https://doi.org/10.1007/s00340-010-4267-4} {\bibfield  {journal} {\bibinfo
  {journal} {Applied Physics B}\ }\textbf {\bibinfo {volume} {101}},\ \bibinfo
  {pages} {705} (\bibinfo {year} {2010})}\BibitemShut {NoStop}%
\bibitem [{\citenamefont {Gopinath}\ \emph {et~al.}(2024)\citenamefont
  {Gopinath}, \citenamefont {Li},\ and\ \citenamefont
  {Davuluri}}]{Gopinath2024}%
  \BibitemOpen
  \bibfield  {author} {\bibinfo {author} {\bibfnamefont {G.}~\bibnamefont
  {Gopinath}}, \bibinfo {author} {\bibfnamefont {Y.}~\bibnamefont {Li}},\ and\
  \bibinfo {author} {\bibfnamefont {S.}~\bibnamefont {Davuluri}},\ }\bibfield
  {title} {\bibinfo {title} {Continuous variable entanglement between
  propagating optical modes using optomechanics},\ }\href
  {https://doi.org/10.1140/epjqt/s40507-024-00252-y} {\bibfield  {journal}
  {\bibinfo  {journal} {EPJ Quantum Technology}\ }\textbf {\bibinfo {volume}
  {11}},\ \bibinfo {pages} {41} (\bibinfo {year} {2024})}\BibitemShut {NoStop}%
\bibitem [{\citenamefont {Pezz\`e}\ \emph {et~al.}(2018)\citenamefont
  {Pezz\`e}, \citenamefont {Smerzi}, \citenamefont {Oberthaler}, \citenamefont
  {Schmied},\ and\ \citenamefont {Treutlein}}]{RevModPhys.90.035005}%
  \BibitemOpen
  \bibfield  {author} {\bibinfo {author} {\bibfnamefont {L.}~\bibnamefont
  {Pezz\`e}}, \bibinfo {author} {\bibfnamefont {A.}~\bibnamefont {Smerzi}},
  \bibinfo {author} {\bibfnamefont {M.~K.}\ \bibnamefont {Oberthaler}},
  \bibinfo {author} {\bibfnamefont {R.}~\bibnamefont {Schmied}},\ and\ \bibinfo
  {author} {\bibfnamefont {P.}~\bibnamefont {Treutlein}},\ }\bibfield  {title}
  {\bibinfo {title} {Quantum metrology with nonclassical states of atomic
  ensembles},\ }\href {https://doi.org/10.1103/RevModPhys.90.035005} {\bibfield
   {journal} {\bibinfo  {journal} {Rev. Mod. Phys.}\ }\textbf {\bibinfo
  {volume} {90}},\ \bibinfo {pages} {035005} (\bibinfo {year}
  {2018})}\BibitemShut {NoStop}%
\bibitem [{\citenamefont {Weinstein}\ \emph {et~al.}(2010)\citenamefont
  {Weinstein}, \citenamefont {Beloy},\ and\ \citenamefont
  {Derevianko}}]{PhysRevA.81.030302}%
  \BibitemOpen
  \bibfield  {author} {\bibinfo {author} {\bibfnamefont {J.~D.}\ \bibnamefont
  {Weinstein}}, \bibinfo {author} {\bibfnamefont {K.}~\bibnamefont {Beloy}},\
  and\ \bibinfo {author} {\bibfnamefont {A.}~\bibnamefont {Derevianko}},\
  }\bibfield  {title} {\bibinfo {title} {Entangling the lattice clock: Towards
  heisenberg-limited timekeeping},\ }\href
  {https://doi.org/10.1103/PhysRevA.81.030302} {\bibfield  {journal} {\bibinfo
  {journal} {Phys. Rev. A}\ }\textbf {\bibinfo {volume} {81}},\ \bibinfo
  {pages} {030302} (\bibinfo {year} {2010})}\BibitemShut {NoStop}%
\bibitem [{\citenamefont {Finkelstein}\ \emph {et~al.}(2023)\citenamefont
  {Finkelstein}, \citenamefont {Bali}, \citenamefont {Firstenberg},\ and\
  \citenamefont {Novikova}}]{Finkelstein_2023}%
  \BibitemOpen
  \bibfield  {author} {\bibinfo {author} {\bibfnamefont {R.}~\bibnamefont
  {Finkelstein}}, \bibinfo {author} {\bibfnamefont {S.}~\bibnamefont {Bali}},
  \bibinfo {author} {\bibfnamefont {O.}~\bibnamefont {Firstenberg}},\ and\
  \bibinfo {author} {\bibfnamefont {I.}~\bibnamefont {Novikova}},\ }\bibfield
  {title} {\bibinfo {title} {A practical guide to electromagnetically induced
  transparency in atomic vapor},\ }\href
  {https://doi.org/10.1088/1367-2630/acbc40} {\bibfield  {journal} {\bibinfo
  {journal} {New Journal of Physics}\ }\textbf {\bibinfo {volume} {25}},\
  \bibinfo {pages} {035001} (\bibinfo {year} {2023})}\BibitemShut {NoStop}%
\bibitem [{\citenamefont {Ha}\ \emph {et~al.}(2021)\citenamefont {Ha},
  \citenamefont {Zhang}, \citenamefont {Dao},\ and\ \citenamefont
  {Overstreet}}]{PhysRevA.103.022826}%
  \BibitemOpen
  \bibfield  {author} {\bibinfo {author} {\bibfnamefont {L.-C.}\ \bibnamefont
  {Ha}}, \bibinfo {author} {\bibfnamefont {X.}~\bibnamefont {Zhang}}, \bibinfo
  {author} {\bibfnamefont {N.}~\bibnamefont {Dao}},\ and\ \bibinfo {author}
  {\bibfnamefont {K.~R.}\ \bibnamefont {Overstreet}},\ }\bibfield  {title}
  {\bibinfo {title} {${D}_{1}$ line broadening and hyperfine frequency shift
  coefficients for $^{87}\mathrm{Rb}$ and $^{133}\mathrm{Cs}$ in ne, ar, and
  ${\mathrm{n}}_{2}$},\ }\href {https://doi.org/10.1103/PhysRevA.103.022826}
  {\bibfield  {journal} {\bibinfo  {journal} {Phys. Rev. A}\ }\textbf {\bibinfo
  {volume} {103}},\ \bibinfo {pages} {022826} (\bibinfo {year}
  {2021})}\BibitemShut {NoStop}%
\bibitem [{\citenamefont {McKinstrie}\ \emph {et~al.}(2021)\citenamefont
  {McKinstrie}, \citenamefont {Stirling},\ and\ \citenamefont
  {Helmy}}]{McKinstrie:21}%
  \BibitemOpen
  \bibfield  {author} {\bibinfo {author} {\bibfnamefont {C.~J.}\ \bibnamefont
  {McKinstrie}}, \bibinfo {author} {\bibfnamefont {T.~J.}\ \bibnamefont
  {Stirling}},\ and\ \bibinfo {author} {\bibfnamefont {A.~S.}\ \bibnamefont
  {Helmy}},\ }\bibfield  {title} {\bibinfo {title} {Laser linewidths:
  tutorial},\ }\href {https://doi.org/10.1364/JOSAB.439882} {\bibfield
  {journal} {\bibinfo  {journal} {J. Opt. Soc. Am. B}\ }\textbf {\bibinfo
  {volume} {38}},\ \bibinfo {pages} {3837} (\bibinfo {year}
  {2021})}\BibitemShut {NoStop}%
\bibitem [{\citenamefont {Scully}\ and\ \citenamefont
  {Zubairy}(1997)}]{Scully_Zubairy_1997}%
  \BibitemOpen
  \bibfield  {author} {\bibinfo {author} {\bibfnamefont {M.~O.}\ \bibnamefont
  {Scully}}\ and\ \bibinfo {author} {\bibfnamefont {M.~S.}\ \bibnamefont
  {Zubairy}},\ }\href@noop {} {\emph {\bibinfo {title} {Quantum Optics}}}\
  (\bibinfo  {publisher} {Cambridge University Press},\ \bibinfo {year}
  {1997})\BibitemShut {NoStop}%
\bibitem [{\citenamefont {McDonnell}\ \emph {et~al.}(2004)\citenamefont
  {McDonnell}, \citenamefont {Stacey},\ and\ \citenamefont
  {Steane}}]{PhysRevA.70.053802}%
  \BibitemOpen
  \bibfield  {author} {\bibinfo {author} {\bibfnamefont {M.~J.}\ \bibnamefont
  {McDonnell}}, \bibinfo {author} {\bibfnamefont {D.~N.}\ \bibnamefont
  {Stacey}},\ and\ \bibinfo {author} {\bibfnamefont {A.~M.}\ \bibnamefont
  {Steane}},\ }\bibfield  {title} {\bibinfo {title} {Laser linewidth effects in
  quantum state discrimination by electromagnetically induced transparency},\
  }\href {https://doi.org/10.1103/PhysRevA.70.053802} {\bibfield  {journal}
  {\bibinfo  {journal} {Phys. Rev. A}\ }\textbf {\bibinfo {volume} {70}},\
  \bibinfo {pages} {053802} (\bibinfo {year} {2004})}\BibitemShut {NoStop}%
\bibitem [{\citenamefont {Yang}\ \emph {et~al.}(2024)\citenamefont {Yang},
  \citenamefont {Wei}, \citenamefont {Xu}, \citenamefont {Wang},\ and\
  \citenamefont {Jin}}]{10330073}%
  \BibitemOpen
  \bibfield  {author} {\bibinfo {author} {\bibfnamefont {L.}~\bibnamefont
  {Yang}}, \bibinfo {author} {\bibfnamefont {B.}~\bibnamefont {Wei}}, \bibinfo
  {author} {\bibfnamefont {H.}~\bibnamefont {Xu}}, \bibinfo {author}
  {\bibfnamefont {Z.}~\bibnamefont {Wang}},\ and\ \bibinfo {author}
  {\bibfnamefont {X.}~\bibnamefont {Jin}},\ }\bibfield  {title} {\bibinfo
  {title} {Linewidth compression of dfb laser with a high ratio based on sil
  and pdh},\ }\href {https://doi.org/10.1109/LPT.2023.3335215} {\bibfield
  {journal} {\bibinfo  {journal} {IEEE Photonics Technology Letters}\ }\textbf
  {\bibinfo {volume} {36}},\ \bibinfo {pages} {23} (\bibinfo {year}
  {2024})}\BibitemShut {NoStop}%
\bibitem [{\citenamefont {Zhang}\ \emph {et~al.}(2003)\citenamefont {Zhang},
  \citenamefont {Wei}, \citenamefont {Xie},\ and\ \citenamefont
  {Peng}}]{Zhang:03}%
  \BibitemOpen
  \bibfield  {author} {\bibinfo {author} {\bibfnamefont {J.}~\bibnamefont
  {Zhang}}, \bibinfo {author} {\bibfnamefont {D.}~\bibnamefont {Wei}}, \bibinfo
  {author} {\bibfnamefont {C.}~\bibnamefont {Xie}},\ and\ \bibinfo {author}
  {\bibfnamefont {K.}~\bibnamefont {Peng}},\ }\bibfield  {title} {\bibinfo
  {title} {Characteristics of absorption and dispersion for rubidium d2 lines
  with the modulation transfer spectrum},\ }\href
  {https://doi.org/10.1364/OE.11.001338} {\bibfield  {journal} {\bibinfo
  {journal} {Opt. Express}\ }\textbf {\bibinfo {volume} {11}},\ \bibinfo
  {pages} {1338} (\bibinfo {year} {2003})}\BibitemShut {NoStop}%
\bibitem [{\citenamefont {Ye}\ \emph {et~al.}(2001)\citenamefont {Ye},
  \citenamefont {Ma},\ and\ \citenamefont {Hall}}]{PhysRevLett.87.270801}%
  \BibitemOpen
  \bibfield  {author} {\bibinfo {author} {\bibfnamefont {J.}~\bibnamefont
  {Ye}}, \bibinfo {author} {\bibfnamefont {L.~S.}\ \bibnamefont {Ma}},\ and\
  \bibinfo {author} {\bibfnamefont {J.~L.}\ \bibnamefont {Hall}},\ }\bibfield
  {title} {\bibinfo {title} {Molecular iodine clock},\ }\href
  {https://doi.org/10.1103/PhysRevLett.87.270801} {\bibfield  {journal}
  {\bibinfo  {journal} {Phys. Rev. Lett.}\ }\textbf {\bibinfo {volume} {87}},\
  \bibinfo {pages} {270801} (\bibinfo {year} {2001})}\BibitemShut {NoStop}%
\bibitem [{\citenamefont {Nevsky}\ \emph {et~al.}(2001)\citenamefont {Nevsky},
  \citenamefont {Holzwarth}, \citenamefont {Reichert}, \citenamefont {Udem},
  \citenamefont {Hänsch}, \citenamefont {Zanthier}, \citenamefont {Walther},
  \citenamefont {Schnatz}, \citenamefont {Riehle}, \citenamefont {Pokasov},
  \citenamefont {Skvortsov},\ and\ \citenamefont {Bagayev}}]{NEVSKY2001263}%
  \BibitemOpen
  \bibfield  {author} {\bibinfo {author} {\bibfnamefont {A.}~\bibnamefont
  {Nevsky}}, \bibinfo {author} {\bibfnamefont {R.}~\bibnamefont {Holzwarth}},
  \bibinfo {author} {\bibfnamefont {J.}~\bibnamefont {Reichert}}, \bibinfo
  {author} {\bibfnamefont {T.}~\bibnamefont {Udem}}, \bibinfo {author}
  {\bibfnamefont {T.}~\bibnamefont {Hänsch}}, \bibinfo {author} {\bibfnamefont
  {J.}~\bibnamefont {Zanthier}}, \bibinfo {author} {\bibfnamefont
  {H.}~\bibnamefont {Walther}}, \bibinfo {author} {\bibfnamefont
  {H.}~\bibnamefont {Schnatz}}, \bibinfo {author} {\bibfnamefont
  {F.}~\bibnamefont {Riehle}}, \bibinfo {author} {\bibfnamefont
  {P.}~\bibnamefont {Pokasov}}, \bibinfo {author} {\bibfnamefont
  {M.}~\bibnamefont {Skvortsov}},\ and\ \bibinfo {author} {\bibfnamefont
  {S.}~\bibnamefont {Bagayev}},\ }\bibfield  {title} {\bibinfo {title}
  {Frequency comparison and absolute frequency measurement of i2-stabilized
  lasers at 532 nm},\ }\href
  {https://doi.org/https://doi.org/10.1016/S0030-4018(01)01190-7} {\bibfield
  {journal} {\bibinfo  {journal} {Optics Communications}\ }\textbf {\bibinfo
  {volume} {192}},\ \bibinfo {pages} {263} (\bibinfo {year}
  {2001})}\BibitemShut {NoStop}%
\bibitem [{\citenamefont {Lurie}\ \emph {et~al.}(2011)\citenamefont {Lurie},
  \citenamefont {Baynes}, \citenamefont {Anstie}, \citenamefont {Light},
  \citenamefont {Benabid}, \citenamefont {Stace},\ and\ \citenamefont
  {Luiten}}]{Lurie:11}%
  \BibitemOpen
  \bibfield  {author} {\bibinfo {author} {\bibfnamefont {A.}~\bibnamefont
  {Lurie}}, \bibinfo {author} {\bibfnamefont {F.~N.}\ \bibnamefont {Baynes}},
  \bibinfo {author} {\bibfnamefont {J.~D.}\ \bibnamefont {Anstie}}, \bibinfo
  {author} {\bibfnamefont {P.~S.}\ \bibnamefont {Light}}, \bibinfo {author}
  {\bibfnamefont {F.}~\bibnamefont {Benabid}}, \bibinfo {author} {\bibfnamefont
  {T.~M.}\ \bibnamefont {Stace}},\ and\ \bibinfo {author} {\bibfnamefont
  {A.~N.}\ \bibnamefont {Luiten}},\ }\bibfield  {title} {\bibinfo {title}
  {High-performance iodine fiber frequency standard},\ }\href
  {https://doi.org/10.1364/OL.36.004776} {\bibfield  {journal} {\bibinfo
  {journal} {Opt. Lett.}\ }\textbf {\bibinfo {volume} {36}},\ \bibinfo {pages}
  {4776} (\bibinfo {year} {2011})}\BibitemShut {NoStop}%
\bibitem [{\citenamefont {Perrella}\ \emph {et~al.}(2013)\citenamefont
  {Perrella}, \citenamefont {Light}, \citenamefont {Anstie}, \citenamefont
  {Baynes}, \citenamefont {Benabid},\ and\ \citenamefont
  {Luiten}}]{Perrella:13}%
  \BibitemOpen
  \bibfield  {author} {\bibinfo {author} {\bibfnamefont {C.}~\bibnamefont
  {Perrella}}, \bibinfo {author} {\bibfnamefont {P.~S.}\ \bibnamefont {Light}},
  \bibinfo {author} {\bibfnamefont {J.~D.}\ \bibnamefont {Anstie}}, \bibinfo
  {author} {\bibfnamefont {F.~N.}\ \bibnamefont {Baynes}}, \bibinfo {author}
  {\bibfnamefont {F.}~\bibnamefont {Benabid}},\ and\ \bibinfo {author}
  {\bibfnamefont {A.~N.}\ \bibnamefont {Luiten}},\ }\bibfield  {title}
  {\bibinfo {title} {Two-color rubidium fiber frequency standard},\ }\href
  {https://doi.org/10.1364/OL.38.002122} {\bibfield  {journal} {\bibinfo
  {journal} {Opt. Lett.}\ }\textbf {\bibinfo {volume} {38}},\ \bibinfo {pages}
  {2122} (\bibinfo {year} {2013})}\BibitemShut {NoStop}%
\end{thebibliography}%


\providecommand{\noopsort}[1]{}\providecommand{\singleletter}[1]{#1}%
%
\end{document}